\documentclass{aa}  
\usepackage[breaklinks,colorlinks,urlcolor=blue,citecolor=blue,linkcolor=blue]{hyperref}
\usepackage{graphicx}
\usepackage{subcaption}
\usepackage{txfonts}
\usepackage{lipsum}
\usepackage{lscape}             
\usepackage{placeins}           

\usepackage{booktabs}
\usepackage{multirow}
\usepackage{threeparttable}
\usepackage{makecell}
\usepackage{geometry}
\usepackage{array}
\usepackage{caption}
\usepackage{adjustbox}

\newcommand{\so}{AT2019wey}
\usepackage{xcolor}
\usepackage{xcolor}          
\usepackage[normalem]{ulem}
\usepackage{xcolor}

\definecolor{pc}{RGB}{0,204,0} 

\definecolor{pcc}{RGB}{204,0,0}        

\definecolor{pccc}{RGB}{102,0,204}   %

\begin{document}

   \title{An atypical X-ray variability component in the black hole candidate AT2019wey}
   \titlerunning{Imaginary quasi-periodic oscillation in AT2019wey}



   \author{Pengcheng Yang\inst{1}\fnmsep\thanks{Corresponding author: pyang@astro.rug.nl}
        \and Mariano M\'endez\inst{1}\fnmsep\thanks{Corresponding author: mariano@astro.rug.nl}
        \and Sandeep K. Rout \inst{2,3}
        \and Candela Bellavita \inst{1,4,5}
        \and Federico Garc\'ia \inst{4,5}
        \and Diego Altamirano \inst{6}
        \and Ole K\"onig \inst{7}
        \and Federico A. Fogantini \inst{4,5}
        }

   \institute{Kapteyn Astronomical Institute, University of Groningen, PO Box 800, NL-9700 AV Groningen, The Netherlands
            \and New York University Abu Dhabi, PO Box 129188, Abu Dhabi, UAE
            \and Center for Astrophysics and Space Science (CASS), New York University Abu Dhabi, PO Box 129188, Abu Dhabi, UAE
            \and Instituto Argentino de Radioastronom\'ia (CCT La Plata, CONICET; CICPBA; UNLP), C.C.5, 1894 Villa Elisa, Argentina
            \and Facultad de Ciencias Astron\'omicas y Geof\'isicas, Universidad Nacional de La Plata, 1900 La Plata, Argentina
            \and School of Physics and Astronomy, University of Southampton, Southampton, Hampshire SO17 1BJ, UK
            \and Center for Astrophysics \textbar \ Harvard \& Smithsonian, 60 Garden Street, Cambridge, MA 02138, USA  
           }

   \date{Received XXXX XX, 2026}


\abstract
{Recent studies have revealed a notable timing feature in several black hole X-ray binaries (BHXBs) during the soft-to-hard transition at the outburst decay. Within a narrow frequency range, the phase lags between high- and low-energy X-ray light curves exhibit a sudden increase, accompanied by a drop in the coherence function. These narrow features have been associated with a quasi-periodic oscillation (QPO) appearing only in the imaginary part of the cross spectrum (CS). This QPO remains undetected in the power density spectrum (PDS) and is known as imaginary QPO.
Motivated by these results, we analyse five years of NICER observations of the BHXB \so\ during its low-hard state (LHS) and hard-intermediate state (HIMS). We find an imaginary QPO in the CS of \so, with similar characteristics as those found in other BHXBs, making \so\ the fifth BHXB in which such QPOs have been found. As the source hardens, the frequency of the imaginary QPO drops from $\sim$~5~Hz to $\sim$~1~Hz, while its phase lag rises from $\sim$~0.3 rad to $\sim$~0.7~rad during the HIMS and from $\sim$~0.5~rad to $\sim$~0.6~rad during the LHS.
During the HIMS, the phase-lag energy spectrum of the imaginary QPO shows a typical U-shaped profile, while the shape changes in the LHS. The rms spectrum of the imaginary QPO rises below $\sim$~2~keV, peaks at around $\sim$~2~keV and decreases at higher energies, which may be associated with the presence of a relatively cool corona. 
We compare the properties of the imaginary QPO with those of the type-B and C QPOs in BHXBs and find a tentative connection to type-C QPOs. Combining the imaginary QPOs detected in \so\ with those reported in other sources, we find a systematic increase of QPO phase lags with QPO frequency. However, we cannot conclude whether the phase lags of imaginary QPOs exhibit the inclination dependence previously observed in type-C QPOs.}


   \maketitle
    \nolinenumbers

\section{Introduction}

Black hole X-ray binaries (BHXBs) consist of a stellar-mass black hole and a companion star.
Transient BHXBs spend most of their lifetimes in quiescence, occasionally undergoing X-ray outbursts that typically last for weeks to months \citep[see e.g.,][for reviews]{2006ARA&A..44...49R, 2023hxga.book..120B}. 
The soft X-ray emission is generally attributed to the accretion disc \citep[][]{1973A&A....24..337S}, which is characterized by a multi-temperature blackbody spectrum \citep[][]{1984PASJ...36..741M, 1986ApJ...308..635M}, peaking at around 0.1–2.5~keV \citep[e.g.][]{1997AIPC..410..141Z,1997ApJ...479L.145B}; a fraction of these soft photons is Compton up-scattered in a plasma of energetic electron \citep[][]{1980A&A....86..121S}, the so-called corona, producing a hard X-ray power-law–like spectrum in the hard X-ray band \citep[e.g.,][]{1997ApJ...479..926M,2007A&ARv..15....1D, 2010LNP...794...17G}.

During an outburst, BHXBs typically trace an anticlockwise `q'-shaped track in the hardness–intensity diagram \citep[HID; e.g.,][]{2001ApJS..132..377H,2004MNRAS.355.1105F,2005A&A...440..207B}, transitioning through four spectral–timing states \citep[see][for a review]{2016ASSL..440...61B}. 
The outburst begins in the low–hard state (LHS), dominated by Comptonised emission, with a weak thermal disc \citep[e.g.,][]{2018MNRAS.481.5560S, 2019ApJ...874..183S}. 
As accretion increases, the source brightens and passes through the hard- and soft–intermediate states (HIMS, SIMS, respectively), where thermal and Comptonised components contribute comparably \citep[e.g.,][]{1997ApJ...479..926M,2019ApJ...874..183S}. As the outburst continues, the source reaches the high–soft state (HSS) and the X-ray spectrum is dominated by the thermal component with a weak power-law component \citep[e.g.,][]{1997ApJ...479..926M, 2007A&ARv..15....1D}.
During the decay of an outburst, the source returns to quiescence through the SIMS, HIMS and LHS. Some outbursts are failed-transition outbursts \citep[][]{2021MNRAS.507.5507A} only stay in the LHS and HIMS.

During outbursts, BHXB X-ray light curves exhibit variability over milliseconds to years \citep[e.g.,][and references therein]{2016AN....337..398M, 2019NewAR..8501524I}. The Fourier power density spectra (PDS) can be modelled with Lorentzians, representing both broadband noise and narrow quasi-periodic oscillations \citep[QPOs; e.g.,][]{1989ARA&A..27..517V, 2002ApJ...572..392B}.
Low-frequency QPOs (LFQPOs) are the most common ones, with frequencies ranging from a few mHz to $\sim$~30 Hz \citep[e.g.,][]{2002ApJ...572..392B, 2006ARA&A..44...49R}.
LFQPOs are further classified into three types: A, B, and C \citep[][]{1999ApJ...526L..33W, 2002ApJ...564..962R, 2004A&A...426..587C, 2005ApJ...629..403C}, based on the spectral state, quality factor\footnote{The quality factor is defined as the ratio of the Lorentzian centroid frequency to its full width at half maximum (FWHM)} ($Q$), the PDS shape and the total fractional root-mean-square (rms) amplitude of the broadband noise and the QPOs. Type-A QPOs are rarely detected, occasionally appearing in the HSS, with centroid frequencies of $\sim6-8$~Hz, low rms (a few percent), and display relatively broad peaks ($Q$ $\lesssim$ 3) \citep[][]{1999ApJ...526L..33W,2004A&A...426..587C,2014SSRv..183...43B}. Type-B QPOs occur only in the SIMS are characterized by frequencies $\lesssim$10 Hz, low rms amplitudes ($\lesssim$ 5\%), narrow peaks ($Q$ $\gtrsim$ 6), and weak ($\lesssim$10\%) red noise \citep[][]{2004A&A...426..587C,2014SSRv..183...43B,2016ASSL..440...61B}. Type-C QPOs typically appear during the LHS and HIMS, spanning a wide range of centroid frequencies ($\sim0.01-30$ Hz), exhibiting high rms amplitude (up to $\sim$~ 20\%) and narrow peaks ($Q$ $\gtrsim$ 8), and are accompanied by strong broadband noise of up to 40\% \citep[][]{2004A&A...426..587C,2011MNRAS.418.2292M, 2014SSRv..183...43B}.

The complex Fourier cross spectrum (CS) and the coherence function provide further X-ray timing insights. The argument of the CS gives the frequency-dependent phase lags between correlated light curves observed simultaneously in two energy bands \citep[][]{1987ApJ...319L..13V, 1999ApJ...510..874N, 2014A&ARv..22...72U}, while the coherence function quantifies, as a function of frequency, the degree of linear correlation between the light curves measured in two energy bands. 
If multiple variability components contribute to a certain frequency range in both energy bands, the coherence function may drop below unity, even if each component individually produces perfectly coherent variability \citep{1997ApJ...474L..43V}. This behaviour applies both to narrow QPOs and broad noise components, and leads to diverse patterns of coherence observed across different sources and observations \citep[e.g.,][]{1999ApJ...510..874N, 2001ApJ...556..515M, 2000ApJ...531L..45C, 2017MNRAS.469.2011R, 2017MNRAS.472.3821R, 2025ApJ...980..251A}.

QPOs are typically identified in the PDS, but sometimes the PDS appears smooth, described by 2-3 broad Lorentzians \citep[e.g.,][]{2024A&A...687A.284K}. 
\citet{2024MNRAS.527.9405M} introduced a novel method for measuring the lags of variability components in X-ray binaries. Their approach involves simultaneously fitting the PDS and the real and imaginary parts of the CS using a linear combination of Lorentzian functions. The model assumes that the individual components are coherent across energy bands but mutually incoherent. This technique enables the detection of weak signals that may be suppressed in the PDS but remain significant in the CS. 
This method has been successfully applied to several BHXBs, including Cygnus~X--1 \citep[][]{2024A&A...687A.284K, 2025A&A...696A.237F}, MAXI~J1820+070 \citep[][]{2024MNRAS.527.9405M, 2025A&A...696A.128B}, and Swift~J1727.8--1613 \citep[][]{2025A&A...703A.257B}. These studies unveiled a new type of  QPO, called `imaginary' QPOs because they are only detected in the imaginary part of the CS. This results in the QPOs having a large phase lag. Therefore, the occurrence of an imaginary QPO is often accompanied by a sharp increase in the phase lags and a drop in coherence function at the QPO centroid frequency. \citet{2024A&A...687A.284K} reported similar narrow features in the phase-lag and coherence-function frequency spectrum in one of the NICER observations of \so, at a frequency where no apparent narrow component was seen in the PDS. Motivated by these findings, here we investigate the presence of imaginary QPOs in \so\ and study their properties in detail.

\so\ was initially discovered as an optical transient by the Asteroid Terrestrial-impact Last Alert System \citep[ATLAS;][]{2018PASP..130f4505T} optical survey in December 2019 \citep[][]{2019TNSTR...3....1T}. X-ray activity was first detected in March 2020 by the Astronomical Roentgen Telescope - X-ray Concentrator \citep[ART-XC;][]{2021A&A...650A..42P} and  the extended Roentgen Survey with an Imaging Telescope Array \citep[eROSITA;][]{2021A&A...647A...1P} on board the Spektrum-Roentgen-Gamma (SRG) satellite \citep[][]{2020ATel13571....1M}. Since then, multi-wavelength observations have been conducted.
\citet{2021ApJ...920..120Y} estimated that the distance to \so\ is between $\sim$~1 and $\sim$~10 kpc and proposed that the compact object was a black-hole candidate from its locations in the $L_{radio}-L_{X}$ \citep[][]{arash_bahramian_2018_1252036} and $L_{opt}-L_{X}$ diagrams \citep[][]{2006MNRAS.371.1334R}. AT2019wey is a low-inclination ($\lesssim$ $30^\circ$) system with a low-mass companion star ($\lesssim$ 1$M_\odot$) and a short orbital period \citep[$\lesssim$~16~h;][]{2021ApJ...920..120Y}. Observations over $\sim$~5 years show that the source is mostly in the LHS and HIMS \citep[][]{2021ApJ...920..121Y}, and may occasionally enter the HSS \citep[][]{2023ATel16197....1N, 2024JHEAp..42..136Y, 2025ATel17190....1R}.

The structure of this paper is as follows: In Sect.~\ref{observation} we describe the NICER observations and data reduction. We give details of the methods and techniques used in the timing analysis in Appendix~\ref{method}. The results, including features in the phase lags and coherence function, and the timing properties of the imaginary QPO, are presented in Sect.~\ref{results}. Finally, in Sect.~\ref{discussion} we discuss our findings.

\begin{figure}[!ht]
\centering
\includegraphics[width=\hsize]{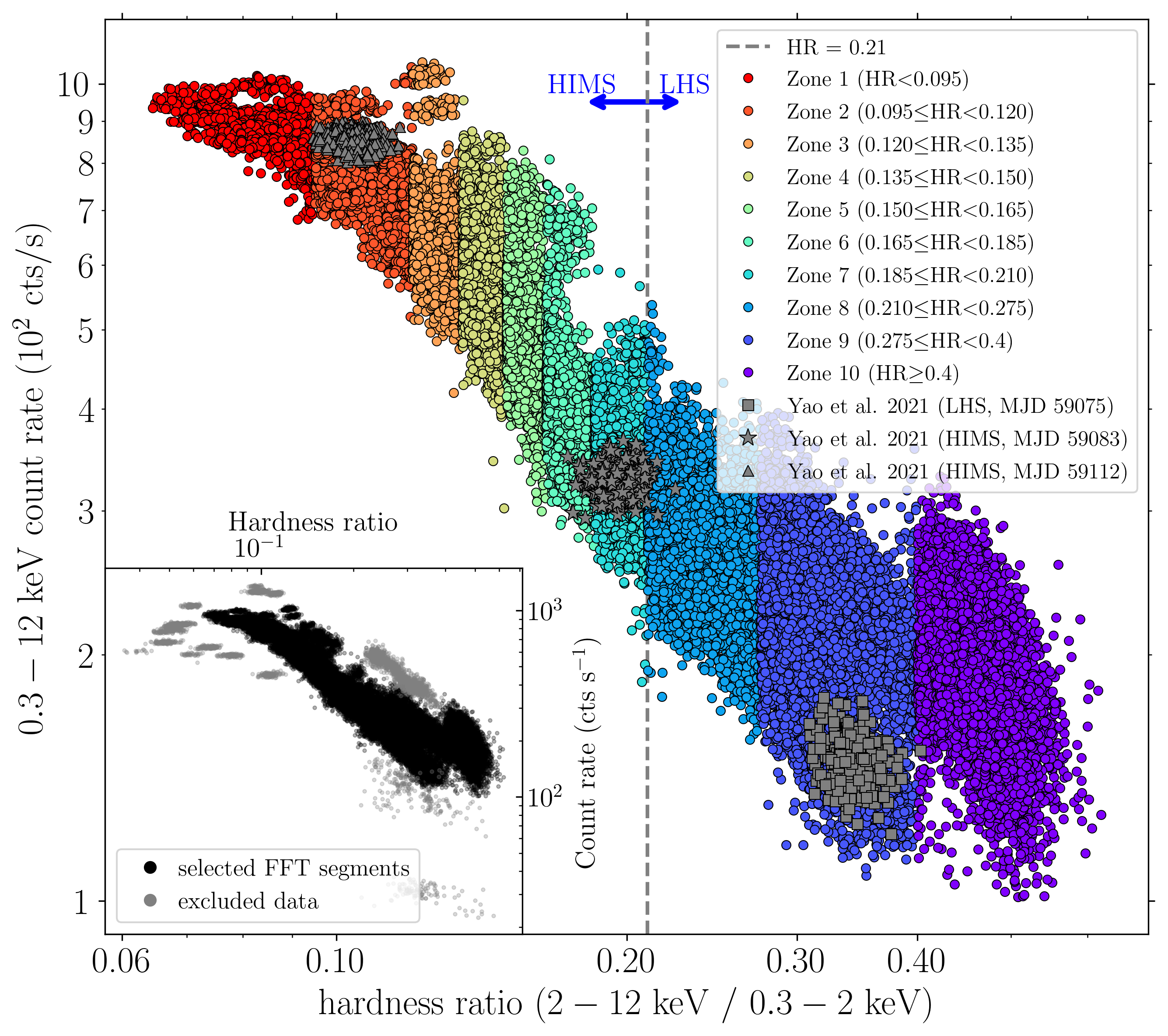}
  \caption{Hardness–intensity diagram (HID) constructed from the NICER observations of \so\ between August 9, 2020 and May 7, 2025. Each data point represents a 16.384~s segment. In the main panel, the ten colour-coded regions correspond to the data groups defined by hardness intervals and used in this work. The grey squares, grey stars, and grey triangles mark the three epochs analysed by \citet{2021ApJ...920..121Y}, identified as the LHS (grey squares) and HIMS (grey stars and triangles), respectively. The dashed vertical line at HR~=~0.21 roughly separates the LHS from the HIMS. In the inset panel, the black points indicate the data used in our analysis (same as in the main panel), while the grey points denote the data that we exclude.}
     \label{fig:group_hid}
\end{figure}

\section{Observation and data reduction}
\label{observation}

Since August 2020, the Neutron Star Interior Composition Explorer \citep[NICER;][]{2016SPIE.9905E..1HG} carried out a long-term monitoring campaign using its onboard X-ray Timing Instrument (XTI). In this work, we make use of all publicly available NICER archival observations of \so\ obtained up to date. The archival dataset comprises more than 600 observations performed between August 4, 2020  and May 7, 2025. We process the data using \texttt{HEASOFT (v6.35.2)} and \texttt{NICERDAS (2025\_06\_11\_V014)}. We use the \texttt{nicerl2} task with CALDB version \texttt{xti20240206} to produce the cleaned event files. We only used orbit night data, which is largely unaffected by the light leak, as reported by the NICER team\footnote{https://heasarc.gsfc.nasa.gov/docs/nicer/analysis\_threads/light-leak-overview/}. To minimise potential instrumental effects that could affect the reliability of the timing analysis, we keep only events with an undershoot rate below $200$~counts/s and an overshoot rate lower than $5$~counts/s, and a cut-off rigidity (COR\_SAX) above $1.5$~GeV/c.  After filtering, over 400 observations remained for further analysis. 

We define several energy bands of interest for extracting light curves and Fourier products. These include broad bands ($0.3-2$~keV, $2-12$~keV and $0.3-12$~keV) as well as narrower sub-bands ($0.3-0.8$~keV, $0.8-1.2$~keV, $1.2-1.6$~keV, $1.6-1.9$~keV, $1.9-2.4$~keV, $2.4-4$~keV, $4-5$~keV and $5-12$~keV).
We extract light curves in the defined energy bands for each observation using the \texttt{nicerl3-lc} task and estimate background light curves with the \texttt{Scorpeon} model. Prior to running \texttt{nicerl3-lc}, we used \texttt{nicerl3-spect} to generate the response files required for background estimation.
We then use the resulting background files to produce background-subtracted light curves and to apply background corrections in the computation of the rms normalisation of the power density and cross spectra.

For the X-ray timing analysis we use GHATS\footnote{https://github.com/ghats-timing/ghats/} to compute the Fast Fourier Transform (FFT) for each energy band of interest of each observation ID, based on the cleaned event files. We compute the FFTs with a segment length of 16.384 s and a time resolution of 0.5 ms. GHATS generates a Leahy-normalized \citep[][]{1983ApJ...266..160L} PDS for each 16.384~s segment of each observation ID.

In this work, one of our objectives is to search for narrow features in the phase lags and coherence function. 
Only a few individual observation IDs of \so\ exhibit narrow features with low signal-to-noise ratio (SNR). 
Previous studies in other sources have shown that the shape of the PDS and CS and their counterparts --- characteristic structures in the phase-lag and coherence-function frequency spectra --- depend on both spectral hardness and source intensity \citep[][]{2025A&A...696A.128B, 2025A&A...696A.237F, 2025ApJ...990...43R}. Based on this, we assume that observations within a narrow hardness ratio interval share similar timing properties.
To enhance the SNR and enable a more detailed investigation of these features, we select 16.384~s segments in a range of HR and average the PDS and real/imaginary parts of CS of these segments.  
After that, using the averaged PDS and real/imaginary parts of CS in two energy bands, we compute the phase-lag frequency spectrum and coherence-function frequency spectrum for each selection.
We correct each PDS for Poisson noise level by subtracting the average power in the 200-800 Hz range where no other timing
features are present.
Following \citet{1990A&A...230..103B}, we normalise the PDS and CS to units of fractional root-mean-square squared per Hz considering the background contribution. We check that the background light curve is consistent with being constant, and therefore adopt the average background count rate of the entire selection in the corresponding energy band for the correction. 
When computing the cross spectrum, phase-lag, and coherence-function frequency spectra, we use the $0.3-2.0$~keV band as the reference band for comparisons between the $0.3-2$~keV and $2-12$~keV ranges. For the energy-dependent analysis, we take the total $0.3-12$~keV band as the reference, while the narrower sub-bands were treated as subject bands.
To correct for partial correlations arising from photons that are both in the subject and reference bands, we subtract the average real part of the cross spectrum measured over a frequency range free of source contributions \citep[][]{2019MNRAS.489.3927I,2024MNRAS.527.7136B,2024MNRAS.527.9405M,2025A&A...696A.128B,2025A&A...696A.237F}.
We rotate\footnote{As explained by \citet{2024MNRAS.527.9405M}, the Levenberg–Marquardt algorithm in XSPEC is more stable when the free parameters are of the same order of magnitude.} all the CS by $45^{\circ}$ to have approximately equal real and imaginary parts; as shown by \citet[][]{2024MNRAS.527.9405M} this is equivalent to a change of coordinates and therefore does not affect the final results.
We applied a logarithmic rebinning to all Fourier products such that the size of each frequency bin increases by $10^{1/100}$ with respect to the previous one to increase the SNR.
For all Fourier products, the segment length and time resolution determine the accessible frequency range, yielding a minimum Fourier frequency of approximately 1/16.384 Hz and a Nyquist frequency of 1000 Hz. Considering the frequency range of the variability components of interest, we restrict our analysis to Fourier products below 20 Hz.

\section{Results}
\label{results}

\begin{figure*}[!ht]
    \centering
    \includegraphics[scale=0.55]{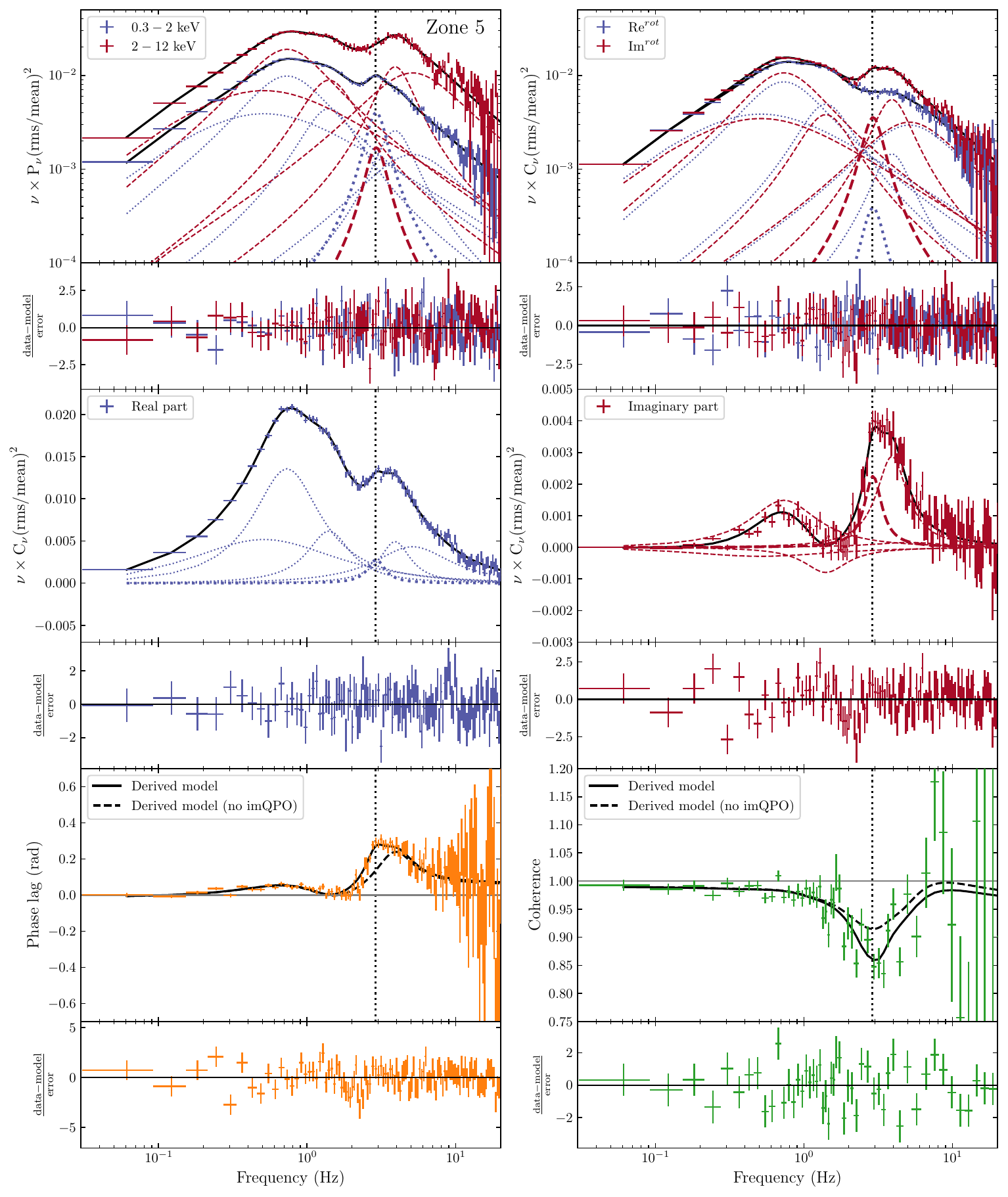}
     \caption{Simultaneous fitting of the PDS and CS of \so\ with a multi-Lorentzian model assuming constant phase lags, using the combined NICER data in Zone 5 as an example.
     Top panels: Best fits (black solid lines) and residuals for the PDS (left panel) in the $0.3-2$~keV (blue) and $2-12$~keV (red) bands, and to the real (blue) and imaginary (red) parts of the CS rotated by $45^{\circ}$ (right panel). 
     Middle panels: The real (left) and imaginary (right) parts of the non-rotated CS, showing a linear scale on the $y$-axis.
     Each model consists of six Lorentzian components; the one corresponding to the imaginary QPO is highlighted with thicker lines.
     Bottom panels: phase-lag (left) and coherence-function (right) frequency spectra, along with their residuals with respect to the model derived (black solid curves) from the simultaneous fitting of PDS and non-rotated CS. The black dashed curves show the derived model without the imaginary QPO and without refitting. We rebinned the plots by a factor of 5 to improve the visibility of the dip in the coherence function}
      \label{fig:example_fit}
\end{figure*}

\begin{table*}[!ht]
\caption[]{\label{tab:qpozones} Best-fit parameters of the imaginary QPO in different zones derived from Lorentzian fits (1$\sigma$ errors).}
\label{table1}
\centering
\resizebox{\textwidth}{!}{
\begin{tabular}{ccccccccccc}
\hline\hline
Zone & $\mathrm{\nu_{imQPO}}$ & FWHM & Q factor & $\mathrm{\Delta\phi}$ & $\mathrm{rms_{imQPO}}$ \tablefootmark{(a)} (\%) & $\mathrm{rms_{imQPO}}$ \tablefootmark{(b)} (\%) &  $\mathrm{rms_{0.1-20~Hz}}$ \tablefootmark{(c)} (\%) & $\mathrm{rms_{0.1-20~Hz}}$ \tablefootmark{(d)} (\%) & $\sqrt{A_\mathrm{{imQPO}}B_\mathrm{{imQPO}}}$ \tablefootmark{(e)} & $C_\mathrm{{imQPO}}$ \tablefootmark{(f)}\\
 & (Hz) & (Hz) &  & (rad) & (0.3-2~keV) & (2-12~keV) & (0.3-2~keV) & (2-12~keV) & & \\ 
\hline
2 & $4.95\pm0.21$ & $3.2\pm0.7$ & $1.5\pm0.3$ & $0.33\pm0.09$ & $3.34\pm0.59$ & $6.10\pm2.00$ & $10.00\pm0.02$ & $23.34\pm0.07$ & $0.0020\pm0.0008$ & $0.0022\pm0.0009$ \\
3 & $4.04\pm0.08$ & $2.5\pm0.3$ & $1.6\pm0.2$ & $0.38\pm0.15$ & $4.89\pm0.41$ & $4.44\pm2.35$ & $13.74\pm0.02$ & $26.34\pm0.07$ & $0.002\pm0.001$ & $0.0025\pm0.0008$ \\
4 & $3.33\pm0.02$ & $1.3\pm0.1$ & $2.5\pm0.2$ & $0.43\pm0.04$ & $4.63\pm0.27$ & $5.94\pm0.85$ & $16.82\pm0.02$ & $28.54\pm0.06$ & $0.0028\pm0.0004$ & $0.0025\pm0.0005$ \\
5 & $2.88\pm0.02$ & $0.9\pm0.1$ & $3.0\pm0.3$ & $0.68\pm0.08$ & $4.45\pm0.29$ & $2.89\pm1.20$ & $19.46\pm0.03$ & $29.68\pm0.07$ & $0.0013\pm0.0005$ & $0.0018\pm0.0004$ \\
6 & $2.48\pm0.02$ & $1.0\pm0.1$ & $2.5\pm0.2$ & $0.77\pm0.07$ & $5.71 \pm 0.34$ & $4.45 \pm 1.01$ & $22.44\pm0.03$ & $30.90\pm0.07$ & $0.0025\pm0.0006$ & $0.0028\pm0.0005$ \\
7 & $2.21\pm0.04$ & $0.7\pm0.1$ & $3.2\pm0.4$ & $0.70\pm0.09$ & $4.53\pm0.48$ & $4.62\pm0.72$ & $25.10\pm0.04$ & $31.13\pm0.07$ & $0.0021\pm0.0004$ & $0.0021\pm0.0005$ \\
\hline
8 & $1.55\pm0.05$ & $0.8\pm0.2$ & $2.1\pm0.7$ & $0.47\pm0.13$ & $5.72\pm1.72$ & $6.22\pm1.95$ & $27.76\pm0.05$ & $30.20\pm0.09$ & $0.004\pm0.002$ & $0.004\pm0.002$ \\
9 & $1.11\pm0.03$ & $0.6\pm0.2$ & $1.9\pm0.5$ & $0.59\pm0.19$ & $4.06\pm1.30$ & $3.52\pm1.70$ & $31.09\pm0.04$ & $28.56\pm0.07$ & $0.001\pm0.001$ & $0.001\pm0.001$ \\
\hline
\end{tabular}
}
\tablefoot{
\tablefoottext{a,b}{The rms amplitude of the imaginary QPO in the, respectively, $0.3-2$~keV and $2-12$~keV bands.}
\tablefoottext{c,d}{The broadband $0.1-20$~Hz rms amplitude in the, respectively, $0.3-2$~keV and $2-12$~keV bands.}
\tablefoottext{e}{$\sqrt{A_\mathrm{{imQPO}}B_\mathrm{{imQPO}}}$, where $A_\mathrm{{imQPO}}$ and $B_\mathrm{{imQPO}}$ are the normalizations of the Lorentzian component representing the imaginary QPO in the $0.3-2$~keV and $2-12$~keV PDS, respectively.}
\tablefoottext{f}{Normalization of the Lorentzian component representing the imaginary QPO in the cross spectrum.}
}

\end{table*}

\begin{figure}[!ht]
    \centering
    \includegraphics[width=\hsize]{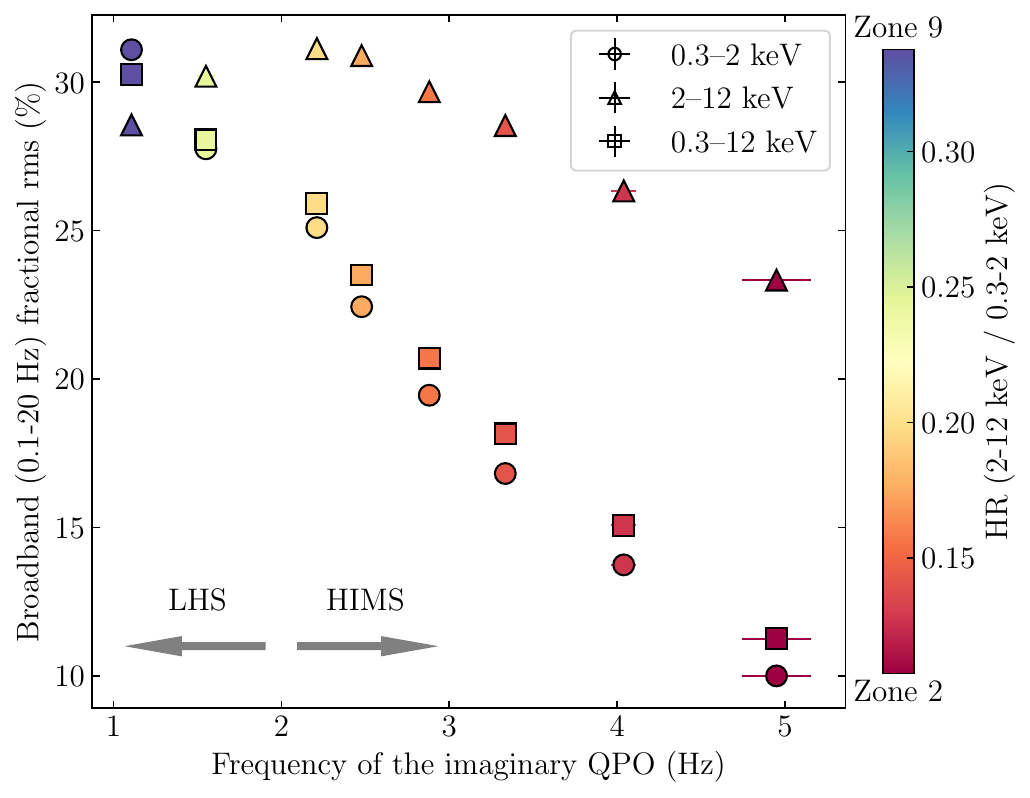}
     \caption{Frequency of the imaginary QPO versus the $0.1-20$ Hz broadband fractional rms in the $0.3-2$~keV (circles), $2-12$~keV (triangles), and $0.3-12$~keV (squares) energy bands derived from our model. Each point represents the observation in a zone. The colour bar indicates the evolution of the hardness ratio across all zones. The two grey arrows separate the observations in the LHS (left) and HIMS (right).}
      \label{fig:nu_brms}
\end{figure}

\begin{figure*}[!ht]
    \centering
    \includegraphics[width=\hsize]{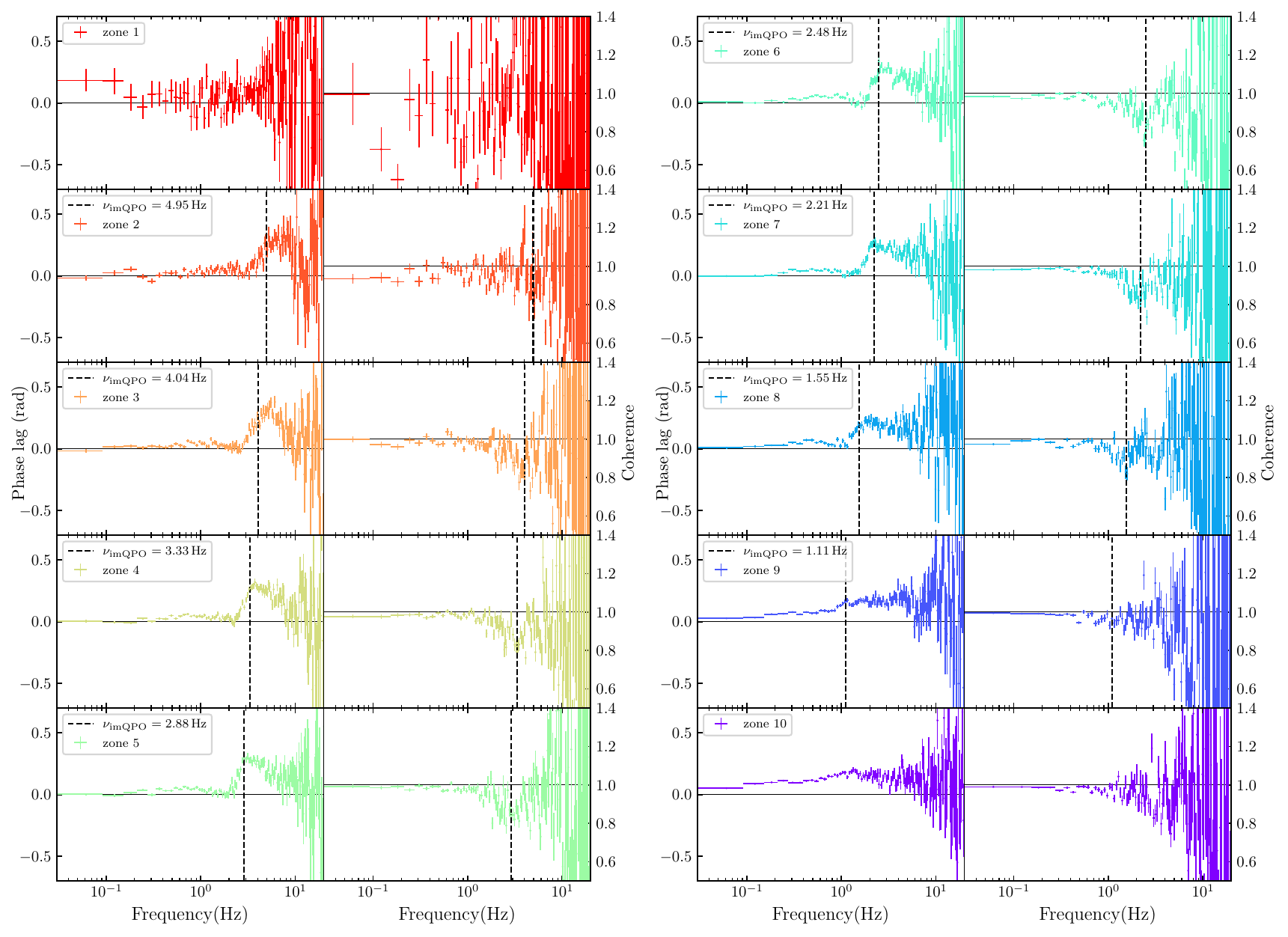}
     \caption{Phase lags and coherence function for each region of the HID of \so\ for the 0.3–2~keV and 2–12~keV bands. The vertical black dashed line marks the frequency of the imaginary QPO (see text) where the phase-lag cliff and coherence-function dip occur. The colour coding used in the frequency spectra for the different zones matches that of Fig.~\ref{fig:group_hid}.}
      \label{fig:cliff_dip}
\end{figure*}

\subsection{Grouping HID}
\label{datagroup}

Fig.~\ref{fig:group_hid} presents the HID of \so, where the hardness ratio (HR) is defined as the ratio of the count rate in the $2-12$~keV and $0.3-2$~keV energy ranges, and the intensity is the count rate in the $0.3-12$~keV energy range. 
Each data point corresponds to a single 16.384 s segment. As mentioned in Sect.~\ref{observation}, only a few observations display possible characteristic narrow timing variability features in the phase-lag and coherence-function frequency spectra extracted from individual observation IDs,  with low SNR.
To ascertain the existence of these features, we combined 16.384~s segments with similar HR to enhance the SNR. 
In the inset of  Fig.~\ref{fig:group_hid} we show in black the points along the main track of the HID for the segments that we retain, and in grey the ones that deviate from the main track and we exclude in the rest of the analysis. 
We first divide the HID into narrow intervals according to hardness ratio width. These intervals allow us to examine the gradual evolution of the PDS and CS shapes, as well as the characteristic features in the phase-lag and coherence-function frequency spectra. Adjacent intervals with similar spectral timing properties are then merged into ten zones, with widths in HR of 0.015 for Zones 1–5, 0.02 for Zone 6, 0.025 for Zone 7, 0.065 for Zone 8, 0.125 for Zone 9, and 0.2 for Zone 10; these zones are colour-coded in Fig.~\ref{fig:group_hid}. Three groups of grey data points represented with squares (MJD 59075), stars (MJD 59083), and triangles (MJD 59112) in the main plot represent the three groups of data from \citet{2021ApJ...920..121Y} during the initial outburst, before MJD 59122. They found that the source starts to enter the HIMS at the position marked by the grey stars. Therefore, we divide the data around the sampling point at HR~=~0.21, as indicated by the vertical black dashed line, to separate the states. Data with HR < 0.21 are classified as belonging to the HIMS, while those with HR~>~0.21 correspond to the LHS. To verify the robustness of the hardness-resolved analysis, we divided the data in each interval shown in Fig.~\ref{fig:group_hid} into high- and low-count-rate subsets, and found no significant differences in their timing properties. We further checked the stability of the PDS and cross-spectral products on shorter timescales by examining them over a number of consecutive NICER orbits within individual observation IDs. We found no significant changes over these shorter timescales.

\subsection{The imaginary QPO}
\label{imaQPO}

We fit simultaneously the two PDS ($0.3-2.0$~keV and $2.0-12$~keV) together with the real and imaginary parts of the CS for each zone, using the method described in Sect.~\ref{fitting}. 
We detect a QPO component in Zones 2--9. 
The best-fitting models for the remaining zones are presented in Appendix~\ref{fig:otherfits}.  We use Zone~5 as a representative example to illustrate our results. As shown in the top panels of Fig.~\ref{fig:example_fit}, the two PDS and the real and imaginary parts of the CS are well described by six Lorentzian components, yielding a reduced $\chi^2$ of 0.969 (410.99 / 424 d.o.f). Both PDS in these energy bands are smooth, showing no narrow peaks that typically indicate the presence of a QPO. However, a narrow Lorentzian component at $\sim$~2.88~Hz is clearly detected in the CS, where it exhibits a much stronger imaginary than real part in the rotated CS, with a quality factor of $Q \sim 3$. In the non-rotated CS, the centroid frequency of this component aligns with the frequency of the peak, as displayed in the middle right panel of Fig.~\ref{fig:example_fit}.
Across Zones 2--9, we find that this component is not always significantly detected in the PDS of either the $0.3-2$ keV or $2-12$ keV energy bands, but it is consistently significant and is strong in the imaginary part of the CS. On this basis, and given that it shares several properties (i.e., the sudden changes in phase lags and coherence function, see the next section) with the imaginary QPOs reported in other sources \citep[e.g.,][]{2024MNRAS.527.9405M,2025A&A...696A.128B,2025A&A...696A.237F,2025A&A...703A.257B}, we will call this component the imaginary QPO.
The bottom panels of Fig.~\ref{fig:example_fit} shows the comparison between the derived model (black solid curve) and the observed phase lags and coherence function, indicating agreement between them.
At $\sim$~2.88 Hz, the phase lags exhibit a sharp increase (`cliff'), accompanied by a simultaneous drop (`dip') in the coherence function. These two behaviours are coincident with the frequency of the imaginary QPO. We note that the black continuous line in the bottom panels is the prediction of the phase lags and coherence function derived from the fits to the two PDS and the real and imaginary parts of the CS, and that we did not fit the phase lags and coherence function. We mark the frequency of the imaginary QPO with a vertical black dotted line in each panel. As illustrated by the black dashed curve in the bottom panel of Fig.~\ref{fig:example_fit}, we also show the model without the imaginary QPO and without refitting. 
This comparison demonstrates that the imaginary QPO is responsible for producing the maximum in the phase lags and the minimum in the coherence function. 
If we refit the spectrum without the imaginary QPO, one of the other components shift and broaden to occupy its frequency range, indicating that the imaginary QPO is a necessary component. However, the resulting fit is significantly worse. Comparing both fits, an F-test yields an F-statistic of 16.85 and a probability of $1.79\times10^{-17}$, indicating that the QPO is highly significant.

Table \ref{table1} lists the parameters (with 1 $\sigma$ errors) of the imaginary QPO in each zone. The frequency of the imaginary QPO decreases from 4.95 Hz to 1.11 Hz as the source spectrum hardens, with the hardness ratio increasing from HR $\approx$ 0.1 (Zone 2) to HR  $\approx$ 0.4 (Zone 9). The quality factor of the imaginary QPO is between $\sim 1.5-3.0$. The phase lags are hard for all zones, increasing from $\sim$~0.3 rad to $\sim$~0.7 rad as the source hardens.
The $\mathrm{\sqrt{A_{i}B_{i}}}$ of the imaginary QPO in each zone is consistent within errors with $\mathrm{C_{i}}$, in accordance with the assumptions described in Sect.~\ref{model}, suggesting that the corresponding Lorentzian components are coherent in the two energy bands.

Motivated by the QPO classification based on the relations between total fractional rms and QPO frequency \citep[][]{2005ApJ...629..403C, 2011MNRAS.418.2292M, 2012MNRAS.427..595M}, 
Fig.~\ref{fig:nu_brms} displays the correlation between the frequency of the imaginary QPO and the broadband ($0.1-20$ Hz) fractional rms in the soft ($0.3-2$~keV), hard ($2-12$~keV) and full ($0.3-12$~keV) energy bands. 
The broadband fractional rms amplitude in the soft band increases from 10\% to 31\% as the QPO frequency decreases from $\sim$~5~Hz to $\sim$~1~Hz and the source moves to the hard state. In the hard band, the broadband fractional rms amplitude increases from 23\% to 31\% as the QPO frequency decreases to $\sim$~2~Hz; below 2~Hz, corresponding to Zones 8 and 9, the rms amplitude starts to decrease as the frequency decreases.
In the full band, the evolution of the broadband rms amplitude with QPO frequency is the same as that in the soft band.

\begin{figure}[!ht]
    \centering
    \includegraphics[width=\hsize]{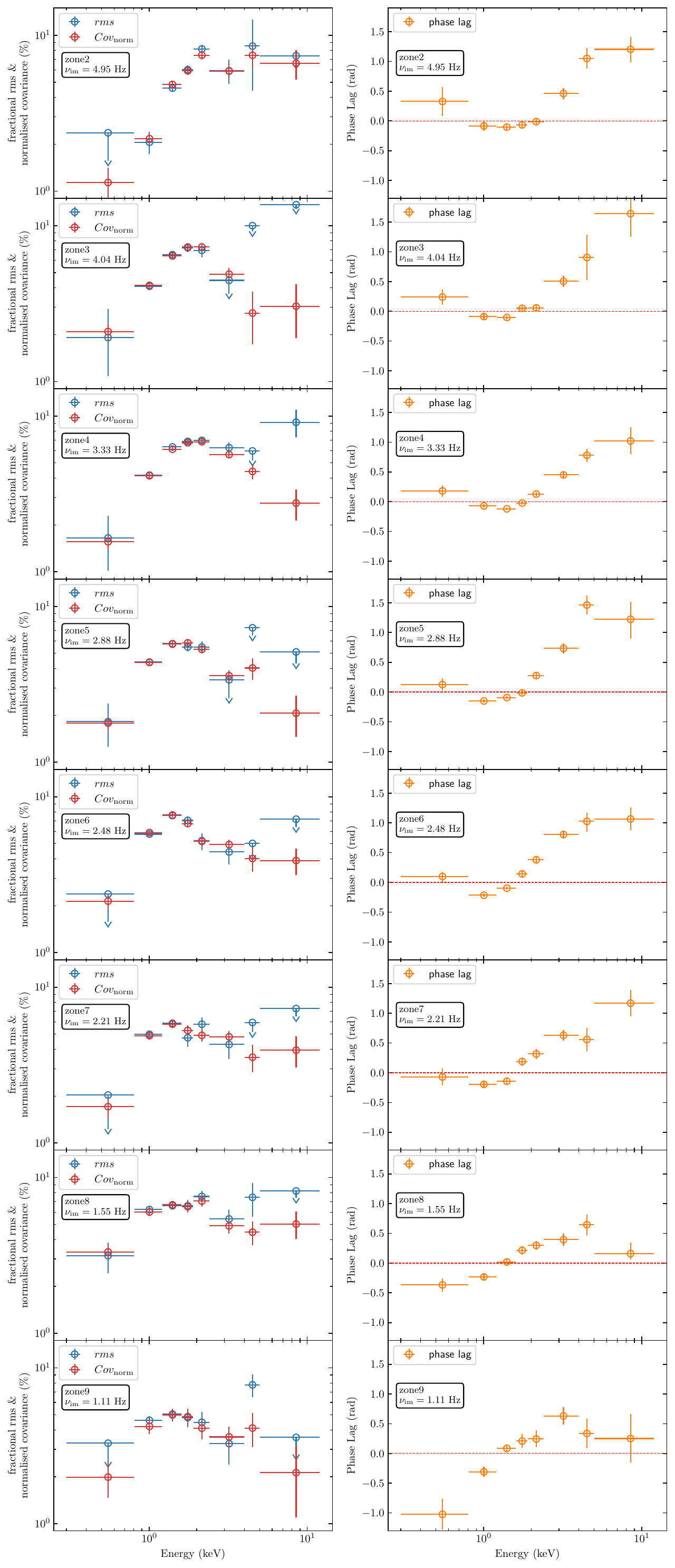}
     \caption{The rms (red), normalised covariance (blue) and phase-lag (orange) spectra of the imaginary QPOs in \so\ for different zone in the HID. All quantities are derived from the fits using our multi-Lorentzian model. The error bars represent 1 $\sigma$ uncertainties. The arrows in the rms spectra represent the 95\% upper limit.}
      \label{fig:enedep}
\end{figure}

\subsection{The increase of the phase lags and the drop of the coherence function}
\label{cliff-dip}
Fig.~\ref{fig:cliff_dip} presents the phase-lag and coherence-function frequency spectra between $0.3-2$~keV and $2-12$~keV for each zone, from the softest zone to the hardest zone. The results for different zones are shown in using the same colours as in Fig.~\ref{fig:group_hid}. The vertical black dashed line indicates the position of the imaginary QPO, obtained from the simultaneous multi-Lorentzian fit to the PDS and CS (see Sect.~\ref{imaQPO} for the imaginary QPO). For Zone 1, there are no apparent characteristic structures in the phase lags and coherence function.
In Zone~2, the phase-lag frequency spectrum exhibits a sudden increase of the phase lag at $\sim$~2~Hz from 0 rad to a maximum value of $\sim$~0.3~rad, followed by a gradual return to 0~rad at $\sim$~10~Hz. We refer to this abrupt increase of the phase lag as the `cliff'. However, no distinct feature is visually identified in the coherence function at this frequency.
From Zone 3 to Zone 8, in addition to the presence of the phase-lag cliff, the coherence exhibits a sudden drop at around the same frequency of the cliff, followed by a recovery to unity, forming a distinct `dip' structure. In Zone 9, a small phase lag cliff appears at $\sim$~1.1 Hz, above which the phase lags form a plateau, while the dip in the coherence function disappears again.
For Zone~10, no obvious phase-lag cliff is observed. Below $\sim$~10~Hz, the phase lags are hard and gradually increases from low frequencies and flattens above $\sim$~1~Hz, whereas the coherence displays no evident structure. In all Zones in which the dip is present, the frequency at which the coherence is minimum coincides with that of the imaginary QPO. We note that, in most cases, the peak of the cliff in the phase lags also corresponds to the imaginary QPO frequency. However, in Zones 2, 3, and 8, the peak appears at a slightly higher frequency, similar to what was observed in Cygnus~X--1 \citep[][]{2025A&A...696A.237F}. In our case, this is due to an additional QPO at an adjacent higher frequency, whose larger phase lag shifts the maximum of the phase-lag frequency spectrum to higher frequency. In Zone 2, the phase lag increases from 0.33 rad at the imaginary QPO frequency to 0.49 rad at the frequency of the adjacent component; in Zone 3, the lags rise from 0.38 rad to 0.81 rad, and in Zone 8, from 0.47 rad to 0.55 rad. In the zones where either the phase-lag cliff or the coherence-function dip is present, we find that both features gradually shift toward lower frequencies as the source hardens.

\subsection{Energy dependence of rms amplitude, normalized covariance and phase lags of the imaginary QPO}
\label{enery_dependence}

To study the energy-dependent properties of the imaginary QPO, we performed separate simultaneous fits, each involving the PDS in the $0.3-12$~keV band, the PDS in one of the narrow bands defined in Sect.~\ref{model}, and the CS between them, following the procedure described in Sect.~\ref{fitting}. For the zones exhibiting the `cliff' or `dip' features, we constructed the fractional rms energy spectra (blue points) of the imaginary QPO measured from the PDS, as presented in Fig.~\ref{fig:enedep}. The error bars indicate 1$\sigma$ uncertainties, while the upper limits of the measurements at higher energies are given at the 95\% confidence level.

Across Zones 2-9, the rms spectra generally show two types of behaviours. Zone 2 displays a rising-flat profile, with the rms increasing at low energies ($\lesssim$ 2.2 keV) and remaining roughly constant at higher energies. In all other zones, the rms spectra exhibit a possible rising-falling profile: the rms increases toward $\sim$~$1.5-2$ keV and then decreases at higher energies, although the detailed shape above $\gtrsim$ 3 keV is often unconstrained. This profile is particularly pronounced in Zone 6. The decreasing trend at higher energies can be identified at 2.2~keV and 3.2~keV. At higher energies the large uncertainties and upper limits make it difficult to assess the shape of the rms spectrum. 
Regarding the evolution of the maximum fractional rms amplitude, during the HIMS (Zones 2-7), the maximum fractional rms amplitude decreases from 7.2\% to 4.3\% as the source hardens. In the LHS (Zones 8–9), the maximum fractional rms amplitude initially rises back to 7.6\% in Zone 8, but then decreases to 5\% in Zone 9 as the hardness increases. Correspondingly, we also find that the energy at which the maximum fractional rms amplitude occurs shifts to lower energies with increasing hardness during HIMS: it is approximately 2.2~keV in Zones 2–4 and about 1.4~keV in Zones 5–7. Similarly, during the LHS, the energy of the maximum fractional rms decreases from 2.2~keV (Zone 8) to 1.4~keV (Zone 9).

To verify the shape of the rms spectra over the entire energy range, we also plot the normalised covariance spectrum (see Appendix.~\ref{define_norcov} for definition), shown as the red points in Fig \ref{fig:enedep}, to cross-check the shape of the rms spectrum at the higher energies.  The normalised covariance is equal to the rms amplitude of the imaginary QPO when the signals in the two energy bands are perfectly coherent.
The normalized covariance spectrum shows good overall agreement with the rms spectrum, particularly below $\sim$~2~keV. Considering the 95\% upper limits, this agreement extends above $\sim$~2~keV as well. In Zone 2, the normalized covariance follows the same shape as the rms spectrum, increasing from $\sim$~1\% at $\sim$~0.55~keV to $\sim$~8\% at $\sim$~2~keV and above. For the other zones, above $\sim$~2~keV, the normalized covariance decreases with increasing energy. We observe that the energy of maximum of the normalised covariance spectrum moves toward lower energies with increasing hardness ratio, from Zone 2 to Zone 9. We conclude that the normalised covariance of the imaginary QPO exhibits a rising-flat trend with energy in Zone 2, whereas in the remaining zones it shows a rising-falling behaviour.

In Fig.~\ref{fig:enedep}, the orange points show the phase-lag spectra. We fit a quadratic polynomial to the data around the minimum of the phase-lag spectra of Zones 2-7, which show the possible U-shape. We find that the energy at which the phase lags are minimum, $E_{min}$, decreases as the source hardens and the frequency of the imaginary QPO decreases, with $E_{min}=$ $1.46 \pm 0.17$~keV (Zone 2), $1.24 \pm 0.09$ (Zone 3), $1.26 \pm 0.05$ (Zone 4), $1.13 \pm 0.06$ (Zone 5), $1.03 \pm 0.04$ (Zone 6), and $0.98 \pm 0.14$ (Zone 7). We fit the $E_{min}-\nu_{imQPO}$ relation with a linear function, $E_{min}=a~\nu_{imQPO}+b$, and find $a=0.18\pm 0.05$ and $b=0.60\pm0.13$, which shows that the trend is 3.6$\sigma$ significant.
For the remaining Zones (8 and 9), the typical U-shaped feature in the phase-lag spectrum disappears as the source hardens further. In Zone 8, the phase lag increases monotonically with energy below $\sim$~4.5~keV and then decreases slightly or remains constant above this energy. In Zone 9, the phase lags show a similar trend to that in Zone 8, but the phase lags remain nearly constant at higher energies within 1$\sigma$ errors.
Based on the state classification in Sect.~\ref{datagroup}, we find that, within the current energy range, the U-shaped phase-lag spectra occur only during the HIMS and disappear in the LHS.

\section{Discussion}
\label{discussion}

We perform a simultaneous analysis of the PDS, CS, phase lags, and coherence function of \so\ using all available NICER observations, following the technique proposed by \citet{2024MNRAS.527.9405M}. We detect a QPO that is not always significant in PDS but is significant in the imaginary part of the CS. Following previous works, we call it imaginary QPO. The phase-lag frequency spectrum exhibits a sudden increase and the coherence-function frequency spectrum shows a drop at the frequency of the QPO. We refer to these two narrow features as the `cliff' and the `dip', respectively.
The fits to the PDS and CS reproduce correctly the cliff of the phase-lag frequency spectrum and the dip of the coherence-function frequency spectrum, lending further support to the assumption that the PDS and CS can be fitted with a linear combination of Lorentzian functions that are coherent in two energy bands but incoherent with one another. 
The imaginary QPO has a large hard phase lag (>~0.3~rad).
We find that the centroid frequency of the imaginary QPO decreases from $\sim$~5 Hz to $\sim$~1 Hz as the source spectrum hardens. The fractional rms amplitude and normalised covariance of the imaginary QPO increase at energies below $\sim$~2~keV but, interestingly, remain more or less constant or show a possible decreasing trend at higher energies. Additionally, we observe a typical U-shaped phase-lag spectrum, and a continuous change in the shape of the phase-lag spectrum as the source transitions from the HIMS to the LHS.

\subsection{The implication of the phase-lag cliff}
\label{discuss_cliff}
At the frequency of the imaginary QPO, the phase lags exhibit a sharp increase, forming a pronounced phase-lag cliff. This behaviour can be naturally attributed to the fact that the imaginary QPO has a larger phase lag than the other components in the same frequency range. 
Several models have been proposed to explain the lags of QPOs. 
In the jet precession model \citep[][]{2021NatAs...5...94M}, if lower-energy photons are emitted from higher altitudes of a precessing jet, soft lag would be produced because the jet base comes into view first. The soft QPO lags predicted by this model are inconsistent with our results.
In the time-dependent Comptonisation model, vKompth \citep[][]{2021MNRAS.503.5522K, 2021MNRAS.501.3173G, 2022MNRAS.515.2099B}, photons from the disc that are inverse-Compton scattered in the corona and escape later will produce hard lags, while reprocessing of returning Comptonised photons (i.e. feedback) in the disc introduces soft lags. This model provides a plausible explanation for the hard phase lags observed in the imaginary QPO.

Interestingly, compared to other sources where the imaginary QPO was detected, the significance of the phase-lag cliff and coherence-function dip in \so\ exhibit a different dependence on the chosen reference energy bands. A clear phase-lag cliff remains visible in \so\ when the $2-4$~keV band is used as the reference band (Fig.~\ref{fig:cliff-dip_above2}), in contrast with other sources \citep[e.g.][]{2024A&A...687A.284K,2025A&A...696A.237F,2025A&A...696A.128B}. 
Assuming the phase lag can still be interpreted within the vKompth model, this finding may indicate that the accretion disc of \so\ contributes non-negligibly to the emission above 2~keV. However, this interpretation may not be unique.
As mentioned in Appendix~\ref{cliff-dip_above2}, the phase-lag cliff above 2~keV is primarily associated with a broad variability component at frequencies higher than that of the imaginary QPO. This broad component has a larger imaginary part than real part, which causes the hard lag. The high-frequency broad variability component in the PDS could be directly associated with the Comptonizing medium \citep{2024A&A...687A.284K}. 
Therefore, the phase-lag cliff above 2~keV could originate solely from the Comptonizing region, in which higher-energy photons undergo more scatterings and are emitted later, naturally producing the hard lag observed between photons above 2~keV.

Under the assumption that the time lag associated with the QPO reflects the light-travel time across the Comptonising region, 
one can describe the evolution of the coronal size. For a 10 $M_{\odot}$ black hole, the size of the corona of \so\ would increases from $\sim3.2\times10^{3}$~km to $1.5\times10^{4}$~km ($215-1023$ $R_{\rm g}$), as the source hardens, corresponding to an increase in the phase lag of the imaginary QPO from $\sim0.3$~rad to $\sim0.7$~rad during the HIMS.
In this case, the inferred values should be regarded as upper limits for a Comptonising cloud with small optical depth ($\tau \gtrsim 0$). The actual size would be much smaller for $\tau \gg 1$.
From the fitting of the energy spectrum, \citet{2021ApJ...920..120Y} found that during the HIMS the inner disc radius remained almost constant at $\sim100$~km (1000~km) assuming a distance of $\sim$~10~kpc (1~kpc). 
The coronal sizes inferred from the time lags are systematically larger than the inner disc radii derived from the spectral fitting. Furthermore, while the spectral analysis suggests that the inner disc radius remains nearly constant during the HIMS, the time lags indicate that the characteristic coronal size increases by approximately an order-of-magnitude.

\subsection{The implication of the coherence-function dip}
\label{discuss_dip}

The coherence can be lower than unity when multiple, physically independent emission regions that are incoherent with one another contribute to the variability in both energy bands, even if each region individually produces perfectly coherent signals \citep{1997ApJ...474L..43V}.
In the PDS of AT2019wey (see Fig.~\ref{fig:example_fit}), multiple Lorentzian components overlap within the narrow frequency range where the coherence dip is observed. If these Lorentzian components correspond to physically independent variability processes, the observed coherence dip is naturally accounted for within the framework proposed by \citet{1997ApJ...474L..43V}.  

\citet[][]{Hua_1997} examined the physical origins of the reduced coherence function in hard X-ray bands above 2~keV. They showed that evolution of the macroscopic parameters of the Comptonising region and/or in the energy of the seed photons during an observation can lead to a loss of coherence. In their scenario, multiple variability components within a given energy band were not considered; instead, the light curves in two different energy bands were assumed to originate from two distinct coronal configurations, effectively representing two independent time series. As a result, the coherence between the two bands naturally drops below unity.
However, in contrast to our findings, such a scenario would lead to a loss of coherence over a broad range of frequencies (see their Fig.~2).

Similar coherence-function dips have been reported during the decay phases of MAXI~J1820+070 \citep[][]{2025A&A...696A.128B}, MAXI J1348--630 \citep[][]{2025ApJ...980..251A}, Swift~J1727.8--1613 \citep[][]{2025A&A...703A.257B}, and Cygnus~X--1 \citep[][]{2025A&A...696A.237F}, all of which appear in the lower branch of the q-shaped HID and hence at relatively low accretion rates \citep[][]{2024A&A...687A.284K}. These results suggest that the sudden coherence dips preferentially occur at low accretion rates, possibly associated with a coronal geometry similar to that inferred for MAXI J1348--630. For \so, fitting the brightest NICER spectrum with \texttt{vphabs*(diskbb + nthComp + relxillCp)} yields a $1.5$–$10$~keV luminosity of $(1.2$–$2.5)\times10^{-2}~L_{\rm Edd}$ for a distance of 10~kpc, or $(1.2$–$2.5)\times10^{-4}~L_{\rm Edd}$ for 1~kpc, assuming a black hole mass of $5$–$10M_\odot$ \citep[][]{2021ApJ...920..120Y}. These values would place \so\ on the lower branch of the HID \citep[][]{2024A&A...687A.284K}.

\subsection{Energy dependence of the normalised covariance and phase lag of the imaginary QPO}
\label{discuss_enedep} 
As described in Sect.~\ref{enery_dependence}, the fractional rms spectra does not provide a good constraint at higher energies, but the covariance spectra give a statistical significant trend with energy. As shown by \citet{2009MNRAS.397..666W}, the normalised covariance is equivalent to the fractional rms amplitude (See Appendix \ref{define_norcov}). Therefore, we will discuss the energy dependence of the variability using the results obtained from the covariance spectra. The covariance spectra of the imaginary QPO in \so\ exhibit two types of profiles: a rising-flat profile observed only in Zone 2, and a rising-falling profile seen in the other zones.
The rising trend of the normalised covariance of the imaginary QPO in \so\ could indicate that the Comptonised component dominates the variability \citep[][]{2020MNRAS.499..851Z}.
Similar rising trend in the fractional rms spectra of the imaginary QPO has been observed in Cygnus~X--1 \citep[][]{2025A&A...696A.237F}, MAXI~J1820+070 \citep[][]{2025A&A...696A.128B}, and Swift~J1727.8--1613 \citep[][]{2025A&A...703A.257B}.

Despite these similarities, \so\ exhibits notable differences compared to these sources. In Cygnus~X--1, MAXI~J1820+070, and Swift~J1727.8--1613, the rms spectra of the imaginary QPOs generally show a rising-flat profile, whereas in \so\ the energy dependence of the normalised covariance predominantly follow a rising-falling trend.
In addition, the energy at which the normalised covariance peaks is systematically lower in \so\ than in MAXI~J1820+070 \citep[][]{2025A&A...696A.128B} and Swift~J1727.8--1613 \citep[][]{2025A&A...703A.257B}.
Specifically, the rms maximum occurs at $\sim1.4$–$2.2$~keV, comparable to that observed in Cygnus~X--1 \citep[][]{2025A&A...696A.237F}, while in MAXI J1820+070 and Swift J1727.8–1613 the fractional rms amplitude continues to increase up to $\sim$~10~keV.

The declining normalised covariance may be associated with the coronal electron temperature. \citet[][see their Fig.~3]{2022MNRAS.515.2099B} showed that in their model a lower coronal electron temperature leads to signs of suppression of variability at higher energies and causes the energy of the maximum of fractional rms amplitude to shift toward lower values. This theoretical prediction appears to be consistent with our observational results, suggesting that AT2019wey hosts a relatively cold corona. Supporting this interpretation, \cite{2026ApJ...998...36S} reported the presence of two Comptonising regions in \so: a hot ($kT_{e}\sim26-370$~keV) and a cold corona ($kT_{e}\sim1.2-2.7$~keV). We therefore suggest that the imaginary QPO in \so\ may originate from variability within the cold corona, which might explain the normalised covariance peaking at low energies and declining toward higher energies.

\citet[][]{2005MNRAS.363.1349G} investigated the pattern of the rms spectra of the broadband noise (BBN) in the BHXBs, XTE~J1650--500 and XTE~J1650--500, over the $2-50$~keV. They found that in the LHS and intermediate state the rms amplitude either decreases or remains approximately constant with increasing energy, while in the HSS it increases with energy. They suggested that a flat rms spectrum arises from variations in the normalisation of the Comptonised component in the energy spectrum, with no change in the spectral shape. They interpreted the decreasing rms amplitude as resulting from the variation in the seed-photon input, together with some required variation in the power released in the Comptonised component.
We note, however, that their analysis concerned the rms spectrum of the BBN component, integrated in the broad frequency band, whereas in this work we focus on the rms spectrum of an individual variability component, the imaginary QPO. It therefore remains unclear whether the idea proposed by \citet[][]{2005MNRAS.363.1349G} for the BBN can also be applied to QPOs. It remains to be seen how the decomposition of the broadband noise into multiple Lorentzian components, each characterised by its own rms spectrum, is reflected in the rms spectrum of BBN.

We note that the decrease of the variability amplitude at higher energies could be due to the partial loss of the imaginary QPO signal in the low signal-to-noise regime. As explained by \citet{2018ApJ...860..167T} for the kHz QPO in the neutron star 4U~0614+09, at high energies the QPO signal can become dominated by noise, leading to a reduction in variability amplitude.

The phase-lag spectra of \so\ exhibit a U-shaped pattern during the HIMS, which disappears in the LHS. A similar behaviour is also observed for the imaginary QPO in MAXI~J1820+070 \citep[see Fig.~B.1 of][]{2025A&A...696A.128B}, where the U-shape in the phase-lag spectrum also starts to disappear as the source hardens, even if in that case the source did not reach the LHS. In both cases, the energy at the minimum of the U-shaped phase-lag spectra shifts toward lower energies as the source spectrum hardens. 
A similar behaviour has been reported in the imaginary QPO in Cygnus~X--1 \citep[][]{2025A&A...696A.237F}, where the energy of the minimum of the phase-lag spectrum was found to scale with the accretion disc temperature. They suggested that this phenomenon can be explained by the time-dependent Comptonisation model, vKompth \citep[][]{2021MNRAS.503.5522K, 2021MNRAS.501.3173G, 2022MNRAS.515.2099B}. In this model, the feedback produces the characteristic U-shaped phase-lag spectrum, with the energy of the phase-lag minimum being positively correlated with the temperature of the soft-photon source (i.e., accretion disc). Qualitatively,
in \so, a similar trend could be seen in Fig.~\ref{fig:enedep} if, as expected in the transition from the HIMS to the LHS, the disc temperature decreases as the source hardens from Zone 2 to Zone 9. The disappearance of the U shape in the LHS might indicate that the minimum of the phase-lag spectrum has moved to even lower energies, below the NICER band. Further investigation of the application of the vKompth model to the data is beyond the scope of this paper and will be addressed in future work.

\subsection{Comparison of imaginary QPOs with type-B/C QPOs}
\label{relation_typ_im}

The rising-flat rms spectrum of the imaginary QPO in \so\ shown in Zone~2 (see Fig.~\ref{fig:enedep}) is reminiscent of that of either the type-C or type-B QPOs in other BHXBs, such as MAXI~J1348--630 \citep[e.g.,][]{2022MNRAS.514.2839A, 2022ApJ...938..108L}, GRS~1915+105 \citep[e.g.,][]{2020MNRAS.494.1375Z, 2021MNRAS.503.5522K}, MAXI~J1535--571 \citep[e.g.,][]{2022MNRAS.512.2686Z}, MAXI~J1820+070 \citep[e.g.,][]{2023MNRAS.525..854M} and GX~339--4 \citep[e.g.,][]{2023MNRAS.519.1336P}.
On the other hand, the rising-falling rms spectrum of the imaginary QPO in \so\ has also been observed for the type-C QPO during the first reflare of MAXI~J1348--630 \citep[][]{2022MNRAS.514.2839A}. 
Moreover, the U-shape phase-lag spectrum of the imaginary QPO in \so\ during the HIMS (Zones 2–7 of Fig.~\ref{fig:enedep}) has often been seen in both type-B and type-C QPOs (see references above).
These comparisons suggest that the energy-dependent rms amplitude and phase-lag characteristics of the imaginary QPO is similar to that of both type-B and type-C QPOs. 

While these similarities point to a close phenomenological connection between the imaginary QPO and type-B/C QPOs, additional criteria are required to identify the specific QPO class.
Given that type-B QPOs are generally observed during the SIMS, whereas the imaginary QPO in \so\ appears in the HIMS and LHS, it is more likely that the imaginary QPO is associated with type-C rather than type-B QPOs. Moreover, during the decay of the outburst in MAXI J1348--630, \citet{2025ApJ...980..251A} reported a narrow Lorentzian component significant both in the PDS and CS, which they identified as a type-C QPO. In that source,  the centroid frequency of the type-C QPO coincides with a cliff-like structure in the phase-lag frequency spectrum and a dip in the coherence function, providing strong evidence that the imaginary QPO in that source, and by extension in \so, is a type-C QPO.

Further support for a type-C association comes from the frequency-dependent timing behaviour.
An anti-correlation between the frequency of the imaginary QPO and the broadband fractional rms amplitude is observed in \so\ (Fig.~\ref{fig:nu_brms}), with the broadband rms reaching relatively high values ($30~\%$). A similar anti-correlation has been observed for the imaginary QPO in MAXI~J1820+070 \citep[see Fig.~6 of][]{2025A&A...696A.128B}. This behaviour closely resembles that of the type-C QPOs observed in GX~339-4 \citep[][]{2011MNRAS.418.2292M}, GRO~J1655--40 \citep[][]{2012MNRAS.427..595M}, and MAXI~J1631--479 \citep[][]{2021MNRAS.505.1213R}.

The frequency of the imaginary QPOs in \so\ decreases with spectral hardening. 
This trend is reminiscent of the type-C QPOs observed during the hard-to-soft transition in BHXBs, whose frequency increases as the energy spectrum softens \citep[e.g.,][]{2000ApJ...531..537S, 2003A&A...397..729V, 2009ApJ...699..453S, 2009ApJ...698.1398M, 2022MNRAS.512.2686Z,2023MNRAS.520..113R}. 
\citet[][]{2021ApJ...920..121Y} reported the detection of the type-C QPO in \so\ during MJD~59112.24--59112.98 and 59083.85--59083.94 (marked as grey triangles and stars in Fig.~\ref{fig:group_hid}), at $6.58 \pm 0.21$~Hz and $2.06 \pm 0.03$~Hz, respectively, whereas the corresponding imaginary QPOs in the same hardness-ratio range are detected at $4.95 \pm 0.21$~Hz and $2.21 \pm 0.04$~Hz. 
The discrepancy between in the second interval of \citet[][]{2021ApJ...920..121Y} and our hardness ratio selection may be explained if the QPO frequency depends both upon hardness ratio and intensity, since in that time interval the source was brighter than in our selection. In the first time interval of \citet[][]{2021ApJ...920..121Y} their QPO frequency and ours are consistent within errors, and the average intensity of the source in their selection and ours are similar. We also examine the data in the time intervals of MJD~59112.24--59112.98 and 59083.85--59083.94, and fit the PDS and CS simultaneously. We detect an imaginary QPO and find that it has the same frequency as that of the type-C QPO detected by \citet[][]{2021ApJ...920..121Y}. Therefore, we suggest that the imaginary QPO in \so\ is the type-C QPO.

\subsection{Possible origin of the imaginary QPO}

The imaginary QPO in \so\ can be interpreted as a distinct signal arising from independent components within the accretion system, as suggested by the linear combination model of multiple Lorentzian functions \citep[][]{2024MNRAS.527.9405M}. 
\citet[][]{2025ApJ...990...43R} discovered a similar coherence dip in Cygnus X--1 at $\sim0.05$~Hz, corresponding to the frequency of a hidden QPO in the hard energy band, in Cygnus~X--1. In their case the dip in the coherence function is present when they use two hard energy bands above 3~keV; they propose that the dip arises from variability produced by Comptonisation in the jet and other Comptonising components. 
In the case of \so, evidence for jet activity has been reported. \citet[][]{2021ApJ...920..120Y} observed a radio brightening as the source transitioned from the LHS to the HIMS, and \citet[][]{2021ApJ...909L..27Y} detected a resolved radio source during the HIMS, which was interpreted as a steady compact jet. These observations raise the possibility that the imaginary QPO in this source may also be linked to jet activity. 
However, in \so\ we do not detect a significant coherence dip and a QPO component when we use only data above 2~keV (see Appendix.~\ref{cliff-dip_above2}).
In addition, the hidden QPO and its associated coherence dip in Cygnus~X--1 reported by \citet[][]{2025ApJ...990...43R} occur at much lower frequencies than those of the imaginary QPO in \so, which may indicate a fundamentally different physical origin from the imaginary QPOs observed at frequencies of $>1$ Hz. Therefore, we cannot conclusively establish a connection between the imaginary QPO in AT2019wey and the jet.

The imaginary QPO could be related to a beat between two broad Lorentzian components \citep{2024A&A...687A.284K}.
However, in \so\ an additional additive narrow Lorentzian component is required to account for the residuals in the imaginary part of the cross spectrum, suggesting that the imaginary QPO is not the result of non-linear coupling between the other broad Lorentzian components present in the PDS. 

The imaginary QPO could also be related to the interplay between the corona and the accretion disc.
In MAXI~J1820+070 \citep[][]{2025A&A...696A.128B} and Cygnus~X--1 \citep[][]{2024A&A...687A.284K,2025A&A...696A.237F}, the imaginary QPO is not detected when reference energy bands above 2~keV are used, and both the associated phase-lag cliff and coherence dip disappear. Similarly, in \so, as mentioned in Appendix.~\ref{cliff-dip_above2}, the imaginary QPO is not significantly detected above 2~keV.
These findings collectively indicate that the soft energy band plays a crucial role in revealing the imaginary QPO, strongly implying that the imaginary QPO may be produced by disc–corona radiative coupling processes \citep[][]{2021MNRAS.503.5522K,2021MNRAS.501.3173G,2022MNRAS.515.2099B,2022A&A...662A.118M}. 
Furthermore, the evolutionary behaviour of the phase-lag spectrum from the HIMS to the LHS, which is consistent with the predictions of the time-dependent Comptonisation model vKompth \citep[][]{2022MNRAS.515.2099B}, as discussed in Sect.\ref{discuss_enedep}, provides additional evidence that disc–corona radiative coupling processes are a plausible mechanism for the production of imaginary QPOs.

\begin{figure}[!ht]
    \centering
    \includegraphics[width=\hsize]{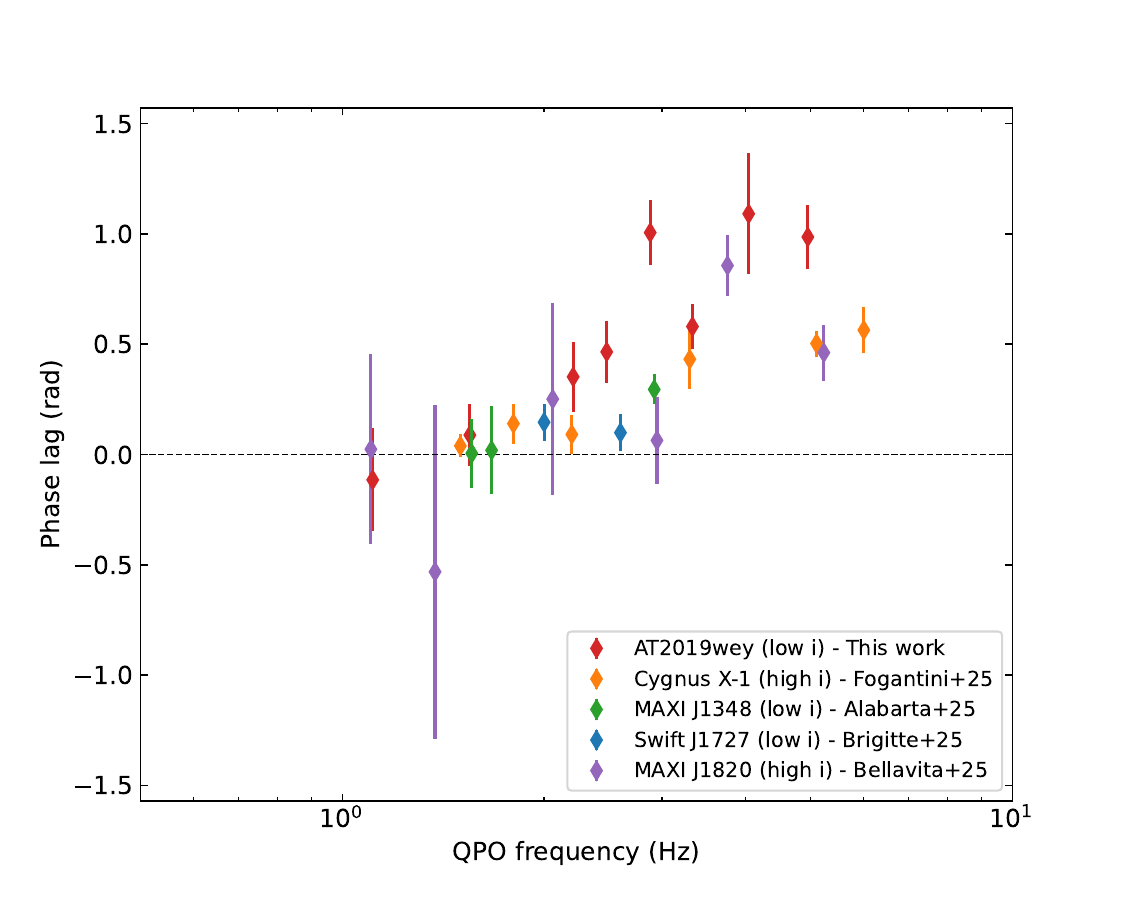}
     \caption{Phase lag of the imaginary QPO as a function of QPO frequency. The inclination (high or low) of each source is inferred from X-ray reflection fitting. The phase lags are computed between the $2-5$~keV (reference band) and $5-12$~keV (subject band) energy bands. For \so, MAXI~J1348--630, MAXI~J1820+070, and Cygnus~X--1, we use NICER data, while for Swift~J1727.8--1613 we use XMM-Newton data.}
      \label{fig_inclination}
\end{figure}

\subsection{Investigation of the inclination dependence of the phase lag of the imaginary QPO}
\label{inclination}

Based on a sample of 15 black hole low-mass X-ray binaries, \citet[][]{2017MNRAS.464.2643V} found that both in low- and high-inclination systems, the type-C QPO exhibit slightly hard lags at low QPO frequencies ($\lesssim$ 1 Hz), whereas at higher frequencies, high-inclination sources transition to soft lags, whereas low-inclination sources show increasingly harder lags. \citet[][]{2017MNRAS.464.2643V} suggested that this behaviour was consistent with the idea that the type-C QPOs reflect the Lense-Thirring precession of the inner accretion flow \citep[][]{2009MNRAS.397L.101I}

We investigate whether the phase lags of the imaginary QPO exhibit a similar inclination dependence to that of the type-C QPOs.
\so\ appears to have a low inclination, as indicated by its single-peaked hydrogen lines in the optical spectrum \citep[][]{2021ApJ...920..120Y}, consistent with the low inclination ($i \lesssim 30^\circ$) inferred from X-ray reflection modelling \citep[][]{2021ApJ...920..121Y}. For Cygnus~X--1, the optical orbital modulation indicates a low orbital inclination of $i \sim 27^\circ$ \citep{2011ApJ...742...84O}, whereas XRISM/Resolve spectral analysis suggests a higher inner disc inclination of $i \sim 63^\circ$ \citep{2025ApJ...995L..12D}. This difference may be explained if the accretion disc is warped \citep[][]{2014ApJ...780...78T}, and the X-ray spectrum reflects a highly inclined disc that is not aligned with the orbit of the secondary in this source. MAXI~J1820+070 has a high orbital inclination of $i=66^\circ-81^\circ$ \citep{2020ApJ...893L..37T} and jet inclination of $i = 64^\circ \pm 5^\circ$ \citep{2021MNRAS.505.3393W}. Relativistic reflection spectroscopy also indicate a high disk inclination of $i = 64^{\circ}{}_{-4}^{+3}$ \citep[][]{2023ApJ...946...19D}.
The jet inclination of Swift~J1727.8–1613 is constrained to be $i < 74^\circ$ \citep{2024ApJ...971L...9W}. A moderate inclination in this source is indicated by other methods, such as estimates from X-ray reflection ($\sim 40^\circ$; \citealt{2023ATel16219....1D}; $\sim 48^\circ$; \citealt{2024ApJ...960L..17P}) and X-ray polarisation observations ($\sim 30$--$60^\circ$; \citealt{2023ApJ...958L..16V}).
Our sample also includes MAXI J1348–630. Although the QPO detected in this source has been classified as type-C, it exhibits properties similar to those of the imaginary QPO \citep[][]{2022MNRAS.514.2839A, 2025ApJ...980..251A}.  MAXI J1348–630 was first identified as a low-inclination ($\sim$30--40$^\circ$) system based on reflection studies \citep[][]{2021MNRAS.508..475C}. \citet[][]{2022MNRAS.511.4826C} measured an inclination of $\sim$29$^\circ$ for the jet of the system with respect to the line of sight.
To ensure as consistent a comparison as possible, we use the common measurements of the inclination from X-ray reflection fitting to define the high ($\gtrsim 60^\circ$) and low ($< 60^\circ$) inclination.
Fig.~\ref{fig_inclination} shows the phase lag of the imaginary QPO between $2-5$~keV and $5-12$~keV energy bands\footnote{Based on the phase-lag spectra reported in the cited references, we compute the weighted average phase lags ($\Delta\phi_{\rm soft}$ and $\Delta\phi_{\rm hard}$) of the imaginary QPO in the $2-5$~keV and $5-12$~keV energy bands, respectively. We then estimate the phase lag between the two bands as $\Delta\phi = \Delta\phi_{hard}-\Delta\phi_{soft}$.} as a function of QPO frequency for all these sources.
This figure shows that the imaginary QPO phase lag above 2~Hz systematically increases with increasing QPO frequency. However, due to the uncertainty in the measurements of the inclination and the limited sample size, we cannot draw a strong conclusion from the current results on whether the phase lag of the imaginary QPO depends on inclination.

\section{Conclusions}
We analysed all available archival NICER observations of \so\ during the LHS and HIMS by the jointly fitting the PDS and CS. We also compare our results with those obtained for other sources and with properties of type-B and type-C QPO. The main conclusions are as follows:
\begin{enumerate}
    \item We detect a QPO signal in \so\ that is not always significant in the power density spectrum but significant in the imaginary part of the cross spectrum. The QPO is always accompanied by a sharp increase (`cliff') in the phase-lag frequency spectrum and a sudden drop (`dip') in the coherence-function frequency spectrum. We refer to this QPO as the imaginary QPO. Our results add a new member to the emerging family of imaginary QPOs. Compared to other sources with imaginary QPOs, we extend the occurrence of this phenomenon to the LHS.
    \item We find a decrease in the normalised covariance of the imaginary QPO above $\sim$ 2~keV, which may be related to a relatively low electron temperature of the Comptonised component in \so.
	\item We find that the phase-lag spectra of the imaginary QPO in \so\ during the HIMS show a U-shaped profile, while the U-shape disappears in the LHS.
    \item The imaginary QPO in \so\ shares several properties with type-C QPOs. We suggest that the imaginary QPO in \so\ could be the type-C QPO.
    \item Whether the phase lags of the imaginary QPOs show an inclination dependence—as seen in type-C QPOs—remains inconclusive based on the current sample size.
\end{enumerate}

While dozens of black-hole X-ray binaries are known, imaginary QPOs have been detected only in a few of them. These QPOs typically appear during the soft-to-hard transition or along the low-accretion-rate branch of the HID. However, their absence in similar transitions of sources like GX 339–4 raises the question of whether this discrepancy may stem from factors like system inclination and disc-corona geometry. Future work should therefore focus on: (1) expanding the sample size of sources showing these QPOs to test for inclination dependence, and (2) systematically comparing disc-corona geometries across different sources to understand why these QPOs occur in some cases but not others.

\begin{acknowledgements} We are grateful to the anonymous reviewer for the comments that helped us improve the manuscript.
FG is a CONICET researcher. FG acknowledges support from PIP 0113 and PIBAA 1275 (CONICET). FG was also supported by grant PID2022-136828NB-C42 funded by the Spanish MCIN/AEI/ 10.13039/501100011033 and “ERDF A way of making Europe”. PY acknowledges support from the China Scholarship Council (CSC), No. 202304910059. PY also thanks Kevin Alabarta for the discussion on the `rising-falling' profile of the rms/covariance spectrum.

\end{acknowledgements}

%
\bibliographystyle{aa}
\bibliography{reference}

\begin{appendix}

\section{Methodology}
\label{method}
\subsection{Multi-Lorentzian model}
\label{model}
We adopt the method proposed by \citet{2024MNRAS.527.9405M} to search for signals in the CS and characteristic structures in the phase-lag frequency spectrum and coherence function. 
This method assumes that the PDS and CS of low-mass X-ray binaries consist of an additive combination of Lorentzian functions, each of which is coherent across different energy bands but incoherent with one another \citep[see also][]{1999ApJ...517..355N}. The technique involves the simultaneous fit of the PDS and the real and imaginary parts of the CS. It has been shown mathematically that this method enhances the detectability of subtle QPOs, which exhibit a higher SNR in the cross spectrum than in the power spectrum. Moreover, the coherence function can reveal signals that are difficult to detect in either the power or cross spectra \citep[e.g.][]{2024MNRAS.527.9405M, 2025A&A...696A.128B,2025ApJ...990...43R}. 

Under the above assumptions, if $x(t)$ and $y(t)$ are correlated light curves in two energy bands, the PDS in the two energy bands can be modelled as:
\begin{equation}
\begin{aligned}
&G_{xx}(\nu) = \sum_{i=1}^{n} G_{xx,i}(\nu) = \sum_{i=1}^{n} A_i L(\nu;\nu_{0,i},\Delta_i)\\
&G_{yy}(\nu) = \sum_{i=1}^{n} G_{yy,i}(\nu) = \sum_{i=1}^{n} B_i L(\nu;\nu_{0,i},\Delta_i),
\label{eq1}
\end{aligned}
\end{equation}
where the $L(\nu;\nu_{0,i},\Delta_i)$ is the Lorentzian functions with centroid frequency $\nu_{0,i}$ and FWHM $\Delta_i$, and $A_i$ and $B_i$ are the normalisations in the respective energy bands.
We take $x$ as the reference band and $y$ as the subject band.
If each Lorentzian is fully coherent in two energy bands, and any two Lorentzians are incoherent, the Real and Imaginary part of the CS between the two light curves can be modelled as a linear combination of Lorentzians:
\begin{equation}
\begin{aligned}
& \mathrm{Re}[G_{xy}(\nu)] = \sum_{i=1}^n C_i\;L(\nu;\nu_{0,i},\Delta_i) \cos{[\Delta\phi_{xy,i}(\nu)+\pi/4]}\\
& \mathrm{Im}[G_{xy}(\nu)] = \sum_{i=1}^n
C_i\;L(\nu;\nu_{0,i},\Delta_i) \sin{[\Delta\phi_{xy,i}(\nu)+\pi/4]},\\
\end{aligned}
\label{eq2}
\end{equation}
where $C_i = \sqrt{A_i B_i}$ and the presumed phase lag between the two signals, $\Delta\phi_{xy,i}(\nu)=g_i(\nu;p_{j,i}$), is frequency-dependent.The additional constant phase shift of $\pi/4$ accounts for the rotation of the cross vector. As explained in \citet[][]{2024MNRAS.527.9405M}, $g_i(\nu;p_{j,i})$ can be any arbitrary function of frequency with parameters $p_{j,i}$ for each Lorentzian; here we take the simplest case of constant phase lag, $g_i(\nu)=2{\pi}k_{i}\in[-\pi, \pi)$, for each Lorentzian, where the $k_i$ are obtained from the fitting of the data.
The total phase lag and the coherence function can be computed from Equation \ref{eq2} \citep[see more details in][]{2024MNRAS.527.9405M}

\subsection{Simultaneous fitting of PDS and CS}
\label{fitting}
Using \texttt{XSPEC 12.15.0f} \citep[][]{1996ASPC..101...17A} we first simultaneously fit the PDS in the $0.3-2$~keV and $2-12$~keV energy bands as well as the rotated Real and Imaginary parts of the CS for the selected 16.384~s segments. To obtain the optimal combination of Lorentzian components, we add them one by one until no significant residuals remain. A Lorentzian is considered to be required in the model if its normalization in either PDS or the CS is at least the $3\sigma$ significant. In Zone 8, shown in the Appendix \ref{fig:otherfits}, the Lorentzian for the imaginary QPO is detected at a 90\% confidence level. During the fitting process, we allow the centroid frequency and  FWHM of each component free to vary, but we link them to be the same in the PDS and the CS. In contrast to $C_i = \sqrt{A_i B_i}$ in the model, in our fitting process we let the normalization of the CS and PDS free to vary independently, and check afterwards whether they are consistent with this relation (Table.~\ref{table1}). Finally, from the best-fit model, we derive the models for the phase lags and the coherence function, and compare them with the observed data. As the PDS is normalized in fractional rms units, integrating the best-fit model over a fixed frequency range yields the total fractional rms. Similarly, the square root of the Lorentzian normalization represents the fractional rms amplitude of the corresponding component.

To investigate the energy-dependent properties of the variability components, from which we derive the fractional rms and phase-lag spectrum, we simultaneously fitted the PDS in the full band (reference band), the PDS in each narrow band (subject bands), and the corresponding cross spectra, using the best-fit model obtained above. In this procedure, we assumed that the centroid frequencies and FWHMs of all Lorentzian components are energy-independent and are therefore linked to be the same.

\subsection{Normalised covariance}
\label{define_norcov}
As described by \citet{2009MNRAS.397..666W} and \citet{2014A&ARv..22...72U}, when two signals are highly coherent over a certain frequency range, the covariance effectively acts as a matched filter, using variability in a high signal-to-noise band to detect much weaker correlated variability in the energy channel of interest. 
The linear multi-Lorentzian model assumes that each Lorentzian function in different energy bands is coherent, so
\begin{equation}
\begin{aligned}
& \gamma^2_{xy, i}(\nu) = \frac{|G_{xy, i}(\nu)|^2}{G_{xx, i}(\nu) G_{yy, i}(\nu)} = \frac{{C_{i}}^2}{A_{i}B_{i}} = 1,\\
\end{aligned}
\label{coherence}
\end{equation}
where the $i$ is for the $i$-th Lorentzian component.
We can define a normalised the covariance of each component following \citet{2009MNRAS.397..666W}:
\begin{equation}
\begin{aligned}
& Cov_{i,norm}  := \sqrt{\frac{|G_{xy, i}(\nu)|^2}{G_{xx, i}(\nu)}} = \frac{{C_{i}}}{\sqrt{A_{i}}}.\\
\end{aligned}
\label{cov}
\end{equation}
Therefore, when the index $i$ corresponds to the imaginary QPO component, the normalised covariance is formally equivalent to the rms of the imaginary QPO, $\sqrt{B_{i}}$. 
On the one hand, $C_{i}$ is the normalisation of the $i$-th Lorentzian in the model for the CS that has better SNR than the PDS. On the other hand, $A_{i}$ is the normalisation of the model for the $0.3-12$~keV reference-band PDS that always provide high SNR.

\section{Remarks on the second imaginary QPO in the CS and weak hard lags below 1~Hz}
\subsection{Second imaginary QPO}
In Fig.~\ref{fig:example_fit}, we identify an additional relatively narrow feature with $Q$~=~1.7 at a higher frequency of 3.75 Hz in Zone~5, located immediately next to the imaginary QPO. We refer to this Lorentzian as the second imaginary QPO. The same component is also visible in some of the other zones (see Appendix \ref{fig:otherfits}). The QPO adjacent to the imaginary QPO that appears as a shoulder of that QPO also shows a relatively large imaginary component, but its centroid frequency does not align with the position of the cliff and dip.
Shoulder-like features also appear in Cygnus~X--1 \citep[][see their Fig.~2]{2025A&A...696A.237F} and MAXI~J1820+070 \citep[][see their Fig.~3]{2025A&A...696A.128B}, although they are not reported. In Swift~J1727.8--1613 \citep[][see their Fig.~2]{2025A&A...703A.257B}, they modelled the PDS and the rotated CS using three Lorentzian components, but the residuals indicate that an additional component is still required at frequencies slightly higher than that of the imaginary QPO.

\subsection{Weak hard lags below 1~Hz}
We also note that in the phase-lag frequency spectrum, the phase lags neither reache zero nor remain constant at frequencies below $\sim$~1~Hz, where weak hard lags are observed. This feature is pronounced in other zones (see Appendix \ref{fig:otherfits}), and moves toward lower frequency as source hardens. 
The phase-lag bump over the low-frequency band is also present in other sources, such as Cygnus~X--1 \citep[][see their Fig.~2]{2025A&A...696A.237F}, MAXI 1820+070 \citep[][see their Fig.~3]{2025A&A...696A.128B} and Swift~J1727.8--1613 \citep[][see their Fig.~2]{2025A&A...703A.257B}.

\section{Best fitting for Zones 2--9}
\label{bestfit_zone2-9}
We simultaneously fit the PDS ($0.3-2$~keV and $2-12$keV) and the CS (real and imaginary parts) of \so\ in Zones 2–9 defined in Sect.\ref{datagroup}, covering the LHS and HIMS. These zones correspond to the intervals where a pronounced cliff in the phase lags and a dip in the coherence function are observed. We applied the technique introduced by \citet[][]{2024MNRAS.527.9405M}, assuming a constant phase-lag model.
Figure~\ref{fig:otherfits} presents the best-fitting PDS in the two energy bands, together with the real and imaginary parts of the rotated CS for each selected zone. The corresponding model predictions for the phase lags and coherence function are also shown. Table \ref{tab:Best-fitting parameters} shows the best-fit parameters for those zones. We note that the frequencies of several common components evolve in the same direction as the imaginary-QPO frequency, indicating that there is correlation between the frequency of the imaginary QPO and the other timing components.

\begin{figure*}[t]
    \centering
    \includegraphics[scale=0.35]{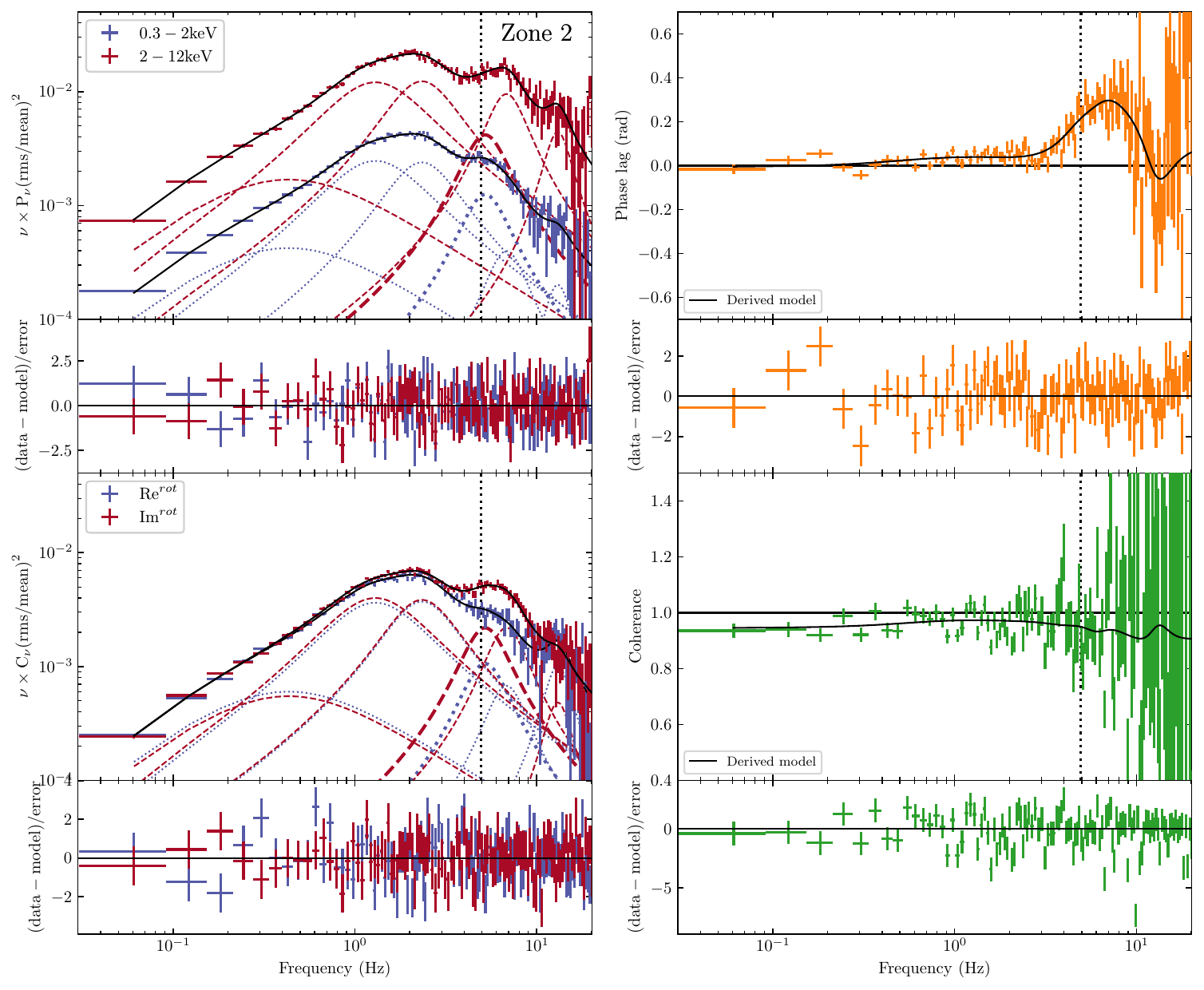}
    \includegraphics[scale=0.35]{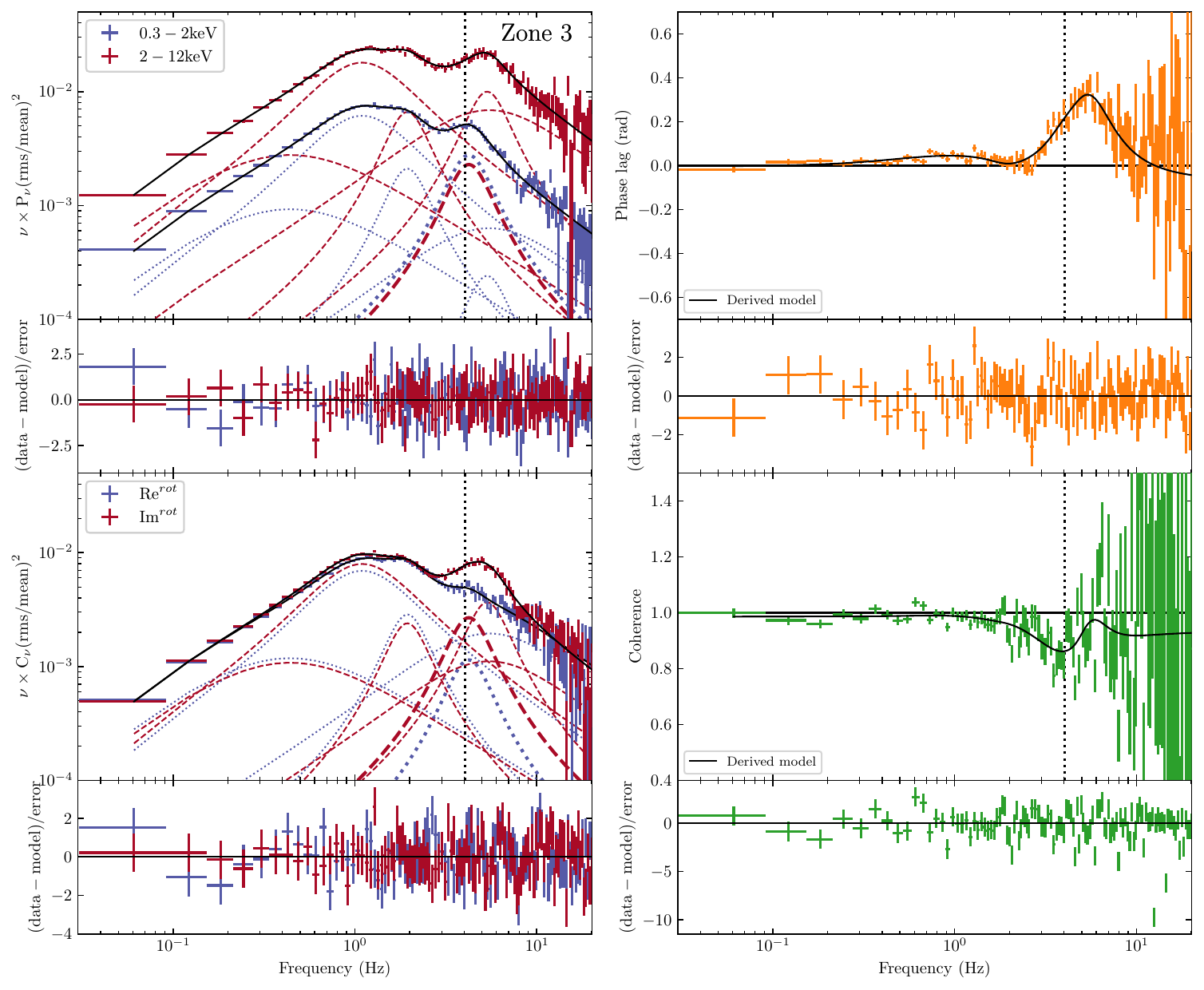}
    \includegraphics[scale=0.35]{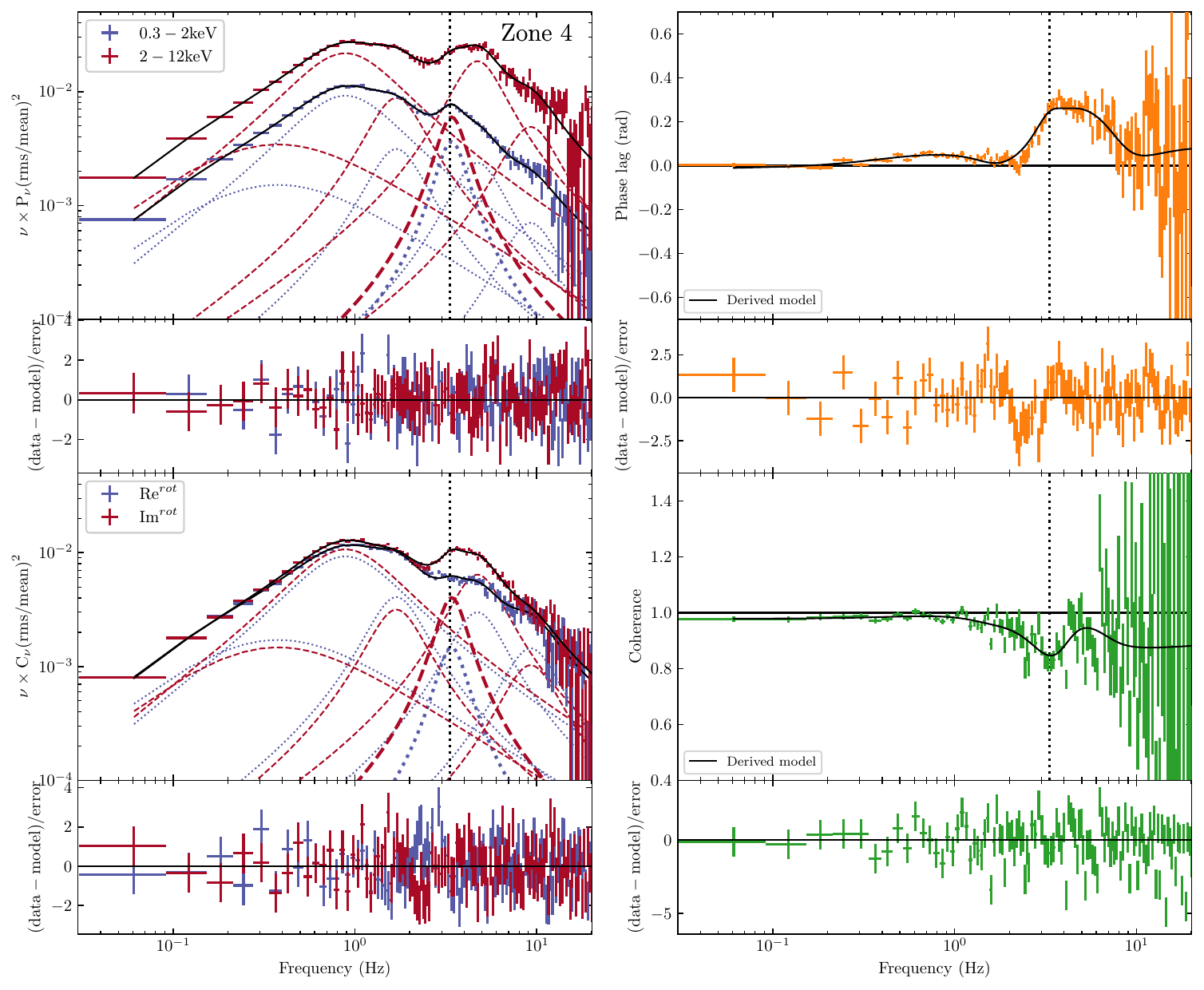}
    \includegraphics[scale=0.35]{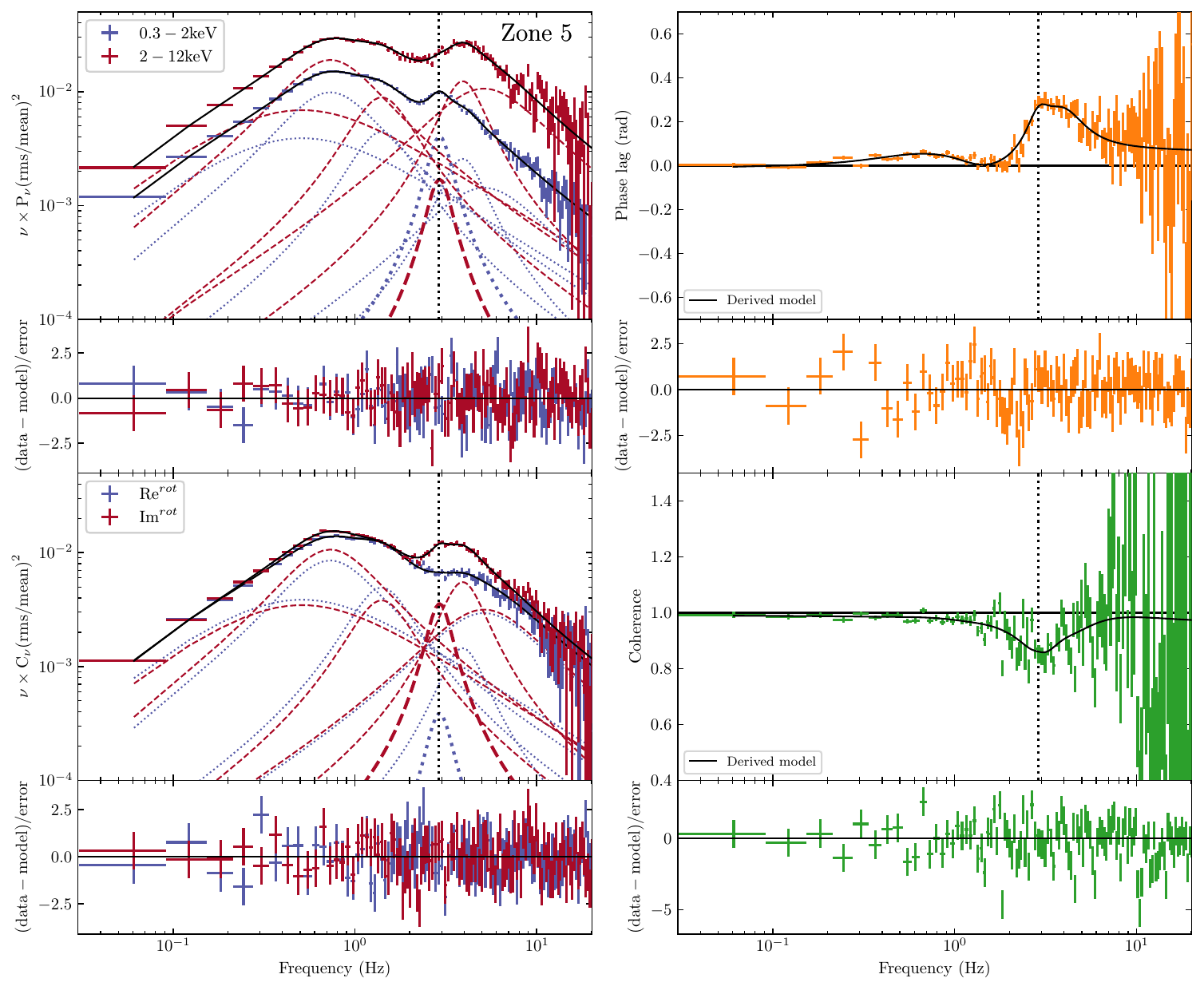}
    \includegraphics[scale=0.35]{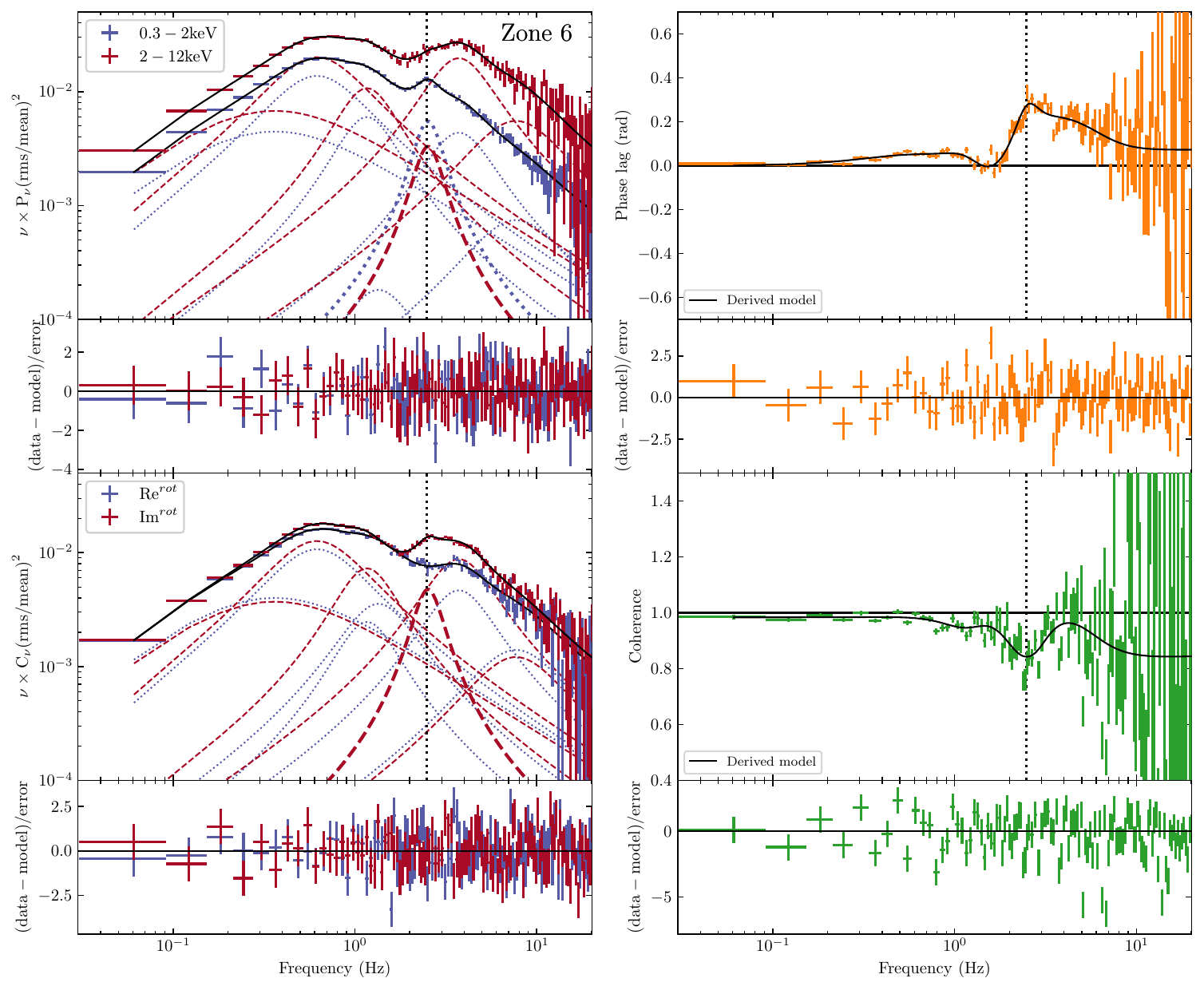}
    \includegraphics[scale=0.35]{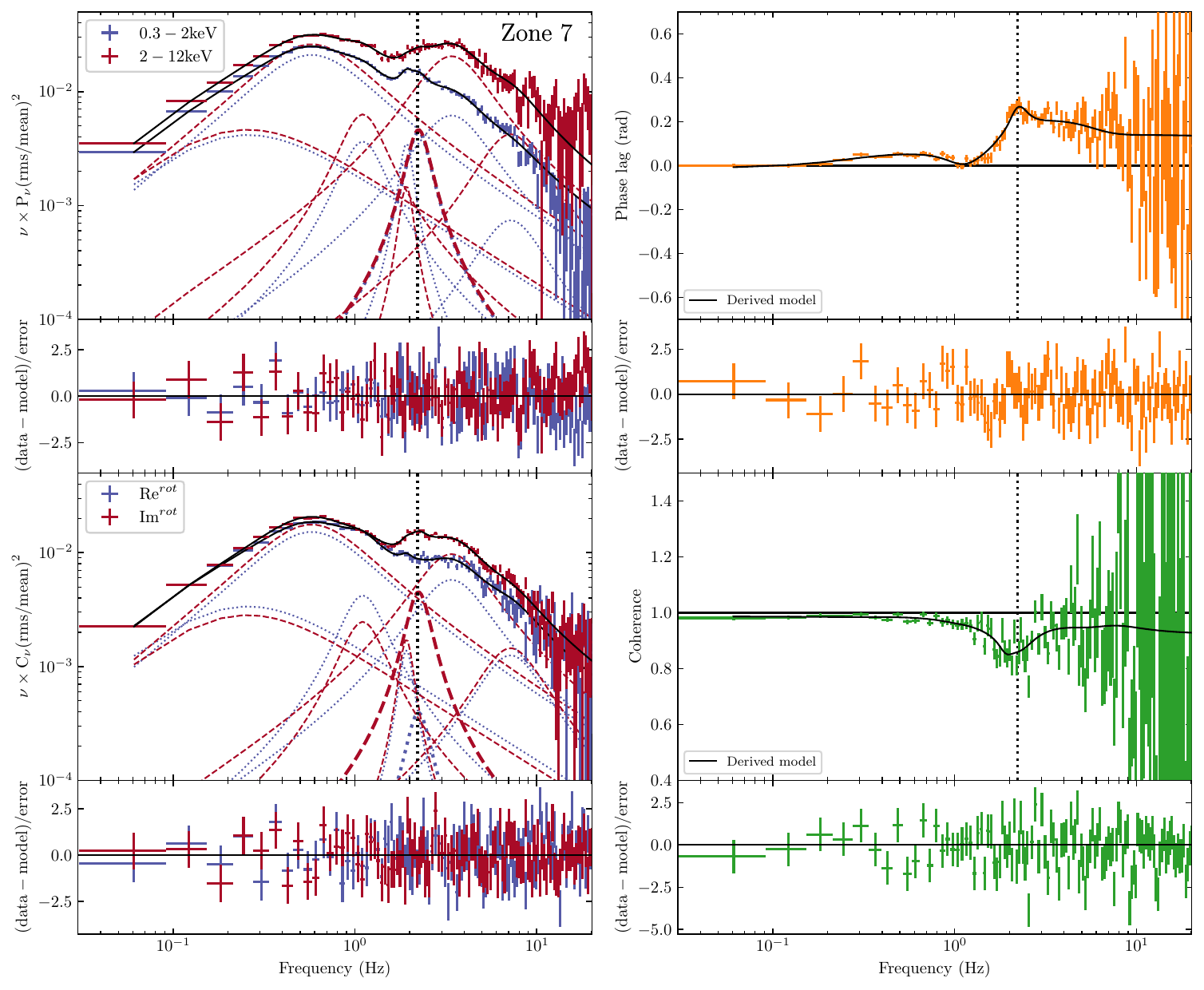}
    
    \caption{The best fitting of the PDS ($0.3-2$~keV and $2-12$~keV) and the corresponding real/imaginary parts of the rotated CS for Zones 2-9,  and the derived model for the phase lags and coherence function.}
    \label{fig:otherfits}
\end{figure*}

\begin{figure*}[t]\ContinuedFloat
    \centering
    \includegraphics[scale=0.35]{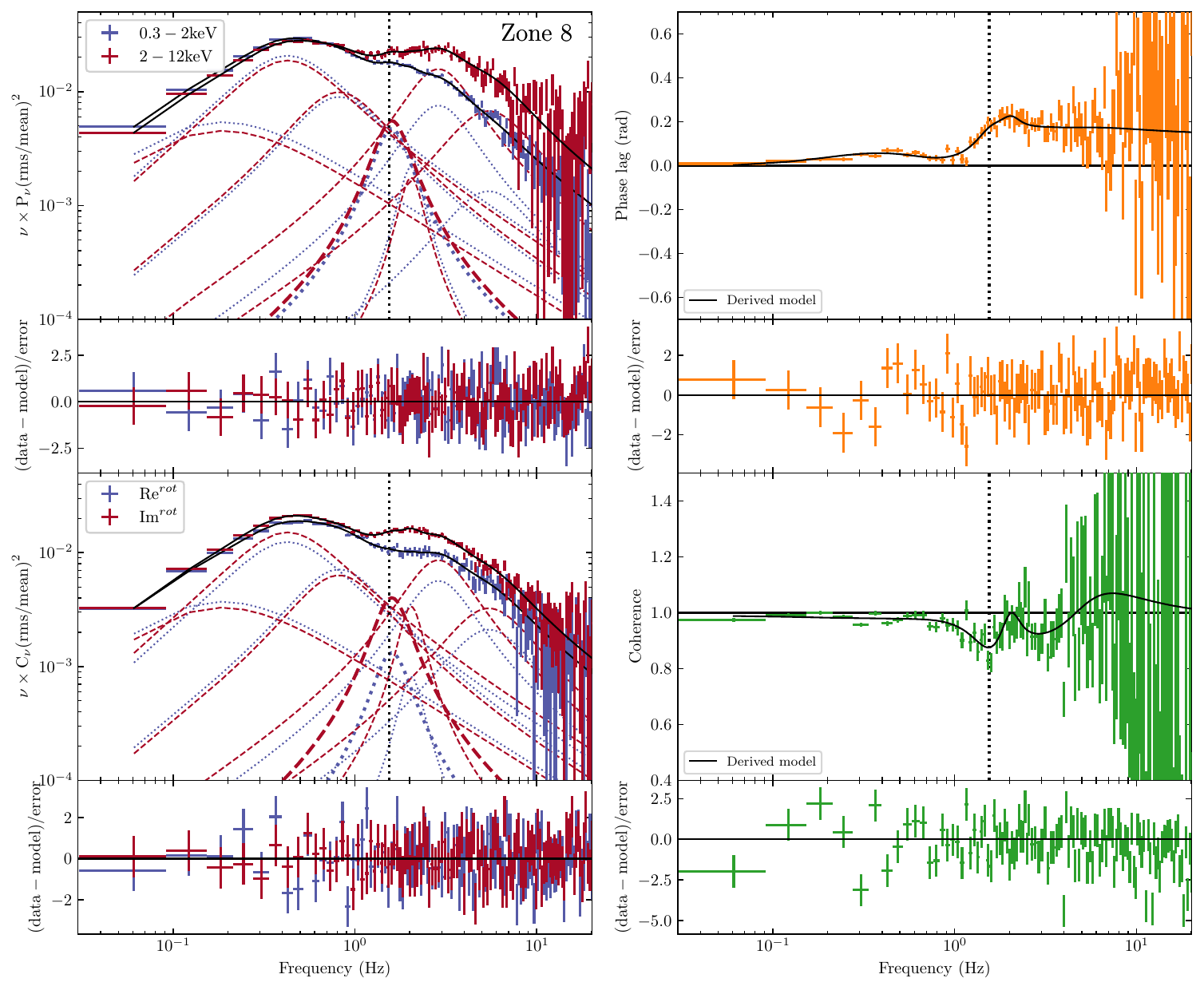}
    \includegraphics[scale=0.35]{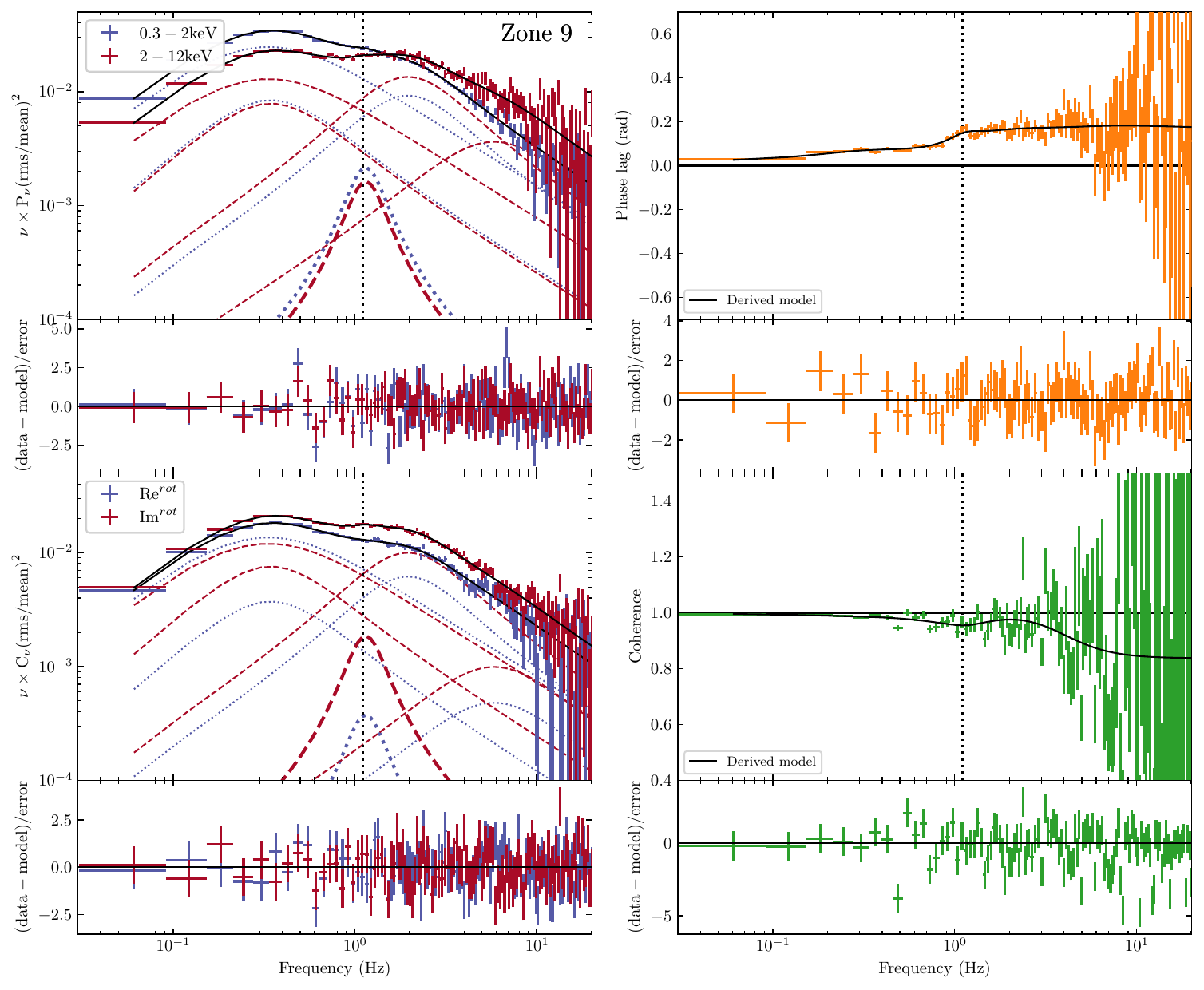}
    \caption{continued.}
\end{figure*}

\begin{table*}[htbp]
\caption{Best-fitting parameters of the combined NICER observations of AT2019wey covering the HIMS and LHS, from Zone 2 to Zone 9, and their corresponding $1\sigma$ uncertainties.}
\label{tab:Best-fitting parameters}
\scriptsize
\setlength{\tabcolsep}{3.5pt}
\renewcommand{\arraystretch}{1.2}
\begin{adjustbox}{max width=\textwidth,center}
\begin{tabular}{lccccccc}
\toprule
Zone & Component & $\nu_0$ (Hz) & FWHM (Hz) & rms$_{0.3-2\ \mathrm{keV}}$ (\%) & rms$_{2-12\ \mathrm{keV}}$ (\%) & Phase lag (rad) & $\chi^2/\mathrm{dof}$ \\
\midrule

\multirow{6}{*}{2}
&  1 & $6.56 \pm 0.33$ & $3.70 \pm 0.54$ & $1.78 \pm 0.83$ & $8.69 \pm 1.45$ & $0.49 \pm 0.09$ & \multirow{6}{*}{430.73/424} \\
&  2 & 1e-10(0.032)    & $0.86 \pm 0.14$ & $3.65 \pm 0.45$ & $7.29 \pm 1.03$ & $-0.05 \pm 0.03$ & \\
&  3$*$ & $4.95 \pm 0.21$ & $3.21 \pm 0.64$ & $3.34 \pm 0.59$ & $6.10 \pm 1.81$ & $0.33 \pm 0.09$ & \\
&  4 & $0.97 \pm 0.02$ & $1.71 \pm 0.24$ & $6.71 \pm 0.98$ & $14.85 \pm 1.95$ & $0.05 \pm 0.01$ & \\
&  5 & $2.06 \pm 0.11$ & $2.22 \pm 0.21$ & $5.65 \pm 0.81$ & $12.82 \pm 1.76$ & $0.02 \pm 0.02$ & \\
&  6 & $12.83 \pm 0.35$ & $4.75 \pm 1.29$ & $1.05 \pm 0.23$ & $4.91 \pm 0.75$ & $-0.34 \pm 0.16$ & \\
\midrule

\multirow{6}{*}{3}
&  1 & $5.11 \pm 0.21$ & $3.10 \pm 0.41$ & 1.43(2.44) & $9.19 \pm 1.87$ & $0.81 \pm 0.11$ & \multirow{6}{*}{406.14/424} \\
&  2 & 1e-10(0.0085) & $0.87 \pm 0.20$ & $5.40 \pm 1.28$ & $9.34 \pm 2.18$ & $-0.04 \pm 0.02$ & \\
&  3$*$ & $4.04 \pm 0.07$ & $2.51 \pm 0.29$ & $4.89 \pm 0.41$ & 4.44(6.32) & $0.38 \pm 0.15$ & \\
&  4 & $0.82 \pm 0.03$ & $1.45 \pm 0.14$ & $10.61 \pm 0.92$ & $18.17 \pm 1.59$ & $0.07 \pm 0.01$ & \\
&  5 & $1.86 \pm 0.04$ & $1.13 \pm 0.14$ & $4.24 \pm 0.71$ & $7.50 \pm 1.35$ & $-0.08 \pm 0.05$ & \\
&  6 & $2.43 \pm 1.46$ & $9.80 \pm 1.27$ & $3.99 \pm 0.52$ & $13.15 \pm 1.04$ & $-0.27 \pm 0.12$ & \\
\midrule
\multirow{6}{*}{4}
&  1 & $4.43 \pm 0.10$ & $3.24 \pm 0.20$ & $4.84 \pm 0.33$ & $13.53 \pm 0.72$ & $0.35 \pm 0.02$ & \multirow{6}{*}{452.78/424} \\
& 2 & 1e-10(0.016) & $0.75 \pm 0.11$ & $6.91 \pm 0.87$ & $10.40 \pm 1.35$ & $-0.07 \pm 0.02$ & \\
& 3$*$ & $3.33 \pm 0.02$ & $1.34 \pm 0.11$ & $4.63 \pm 0.27$ & $5.94 \pm 0.85$ & $0.43 \pm 0.04$ & \\
& 4 & $0.66 \pm 0.02$ & $1.18 \pm 0.07$ & $13.03 \pm 0.67$ & $20.00 \pm 1.04$ & $0.07 \pm 0.01$& \\
& 5 & $1.58 \pm 0.04$ & $1.21 \pm 0.10$ & $5.68 \pm 0.59$ & $9.52 \pm 0.91$ & $-0.13 \pm 0.03$ & \\
& 6 & $8.64 \pm 0.49$ & $6.61 \pm 0.95$ & $2.69 \pm 0.30$ & $7.09 \pm 0.93$ & $ -0.17 \pm 0.08$ & \\
\midrule
\multirow{6}{*}{5}
& 1 & $3.75 \pm 0.08$  & $2.26 \pm 0.25$ & $4.65 \pm 0.63$ & $10.18 \pm 1.68$ & $0.53 \pm 0.09$ & \multirow{6}{*}{410.99/424} \\
& 2 & 1e-10(0.0.042) & $1.02 \pm 0.22$ & $11.07 \pm 1.72$ & $14.70 \pm 2.50$ & $-0.05 \pm 0.02$ & \\
& 3$*$ & $2.88 \pm 0.21$ & $0.95 \pm 0.09$ & $4.45 \pm 0.28$ & $2.89 \pm 1.19$ & $0.68 \pm 0.07$ & \\
& 4 & $0.58 \pm 0.02$ & $0.88 \pm 0.07$ & $12.88 \pm 1.37$ & $17.87 \pm 1.84$ & $0.11 \pm 0.02$ & \\
& 5 & $1.28 \pm 0.04$ & $1.13 \pm 0.09$ & $7.14 \pm 0.81$ & $10.12 \pm 1.24$ & $-0.13 \pm  0.03$ & \\
& 6 & $3.75 \pm 0.49$ & $7.00 \pm 0.76$ & $5.19 \pm 0.75$ & $14.14 \pm 1.63$ & $0.03 \pm 0.05$ & \\
\midrule
\multirow{6}{*}{6}
& 1 & $3.35 \pm 0.09$ & $3.27 \pm 0.24$ & $7.41 \pm 0.54$ & $15.63 \pm 1.15$ & $0.29 \pm 0.03$ & \multirow{7}{*}{440.63/418} \\
& 2 & $0.03 \pm 0.02$ & $0.61 \pm 0.11$ & $10.99 \pm 2.08$ & $13.55 \pm 2.54$ & $-0.04 \pm 0.02$ & \\
& 3$*$ & $2.47 \pm 0.01$ & $1.00 \pm 0.08$ & $5.71 \pm 0.34$ & $4.45 \pm 1.01$ & $0.77 \pm 0.07$ & \\
& 4 & $0.47 \pm 0.02$ & $0.77 \pm 0.05$ & $15.77 \pm 1.39$ & $19.05\pm 1.73$ & $0.08 \pm 0.03$ & \\
& 5 & $1.06 \pm 0.03$ & $0.94 \pm 0.07$ & $8.59 \pm 1.16$ & $11.51 \pm 2.91$ & $0.51 \pm 0.54$ & \\
& 6 & $6.05 \pm 1.33$ & $9.33 \pm 1.09$ & $3.69 \pm 1.21$ & $9.75 \pm 2.48$ & $-0.07 \pm 0.11$ & \\
& 7 & $1.16 \pm 0.22$ & $1.28 \pm 0.24$ & 1.89(8.36) & 0.001(8.83) & $-1.30 \pm 0.87$& \\
\midrule
\multirow{7}{*}{7}
&  1 & $3.00 \pm 0.07$ & $3.16 \pm 0.18$ & $8.99 \pm 0.38$ & $16.37 \pm 0.70$ & $0.25 \pm 0.02$ & \multirow{6}{*}{471.63/418} \\
& 2 & $0.03 \pm 0.02$ & $0.50 \pm 0.12$ & $11.20 \pm 2.07$ & $11.81 \pm 2.34$ & $ -0.09 \pm 0.03$ & \\
& 3$*$ & $2.21 \pm 0.04$ & $0.68 \pm 0.09$ & $4.52 \pm 0.48$ & $4.61 \pm 0.72$ & $0.70 \pm 0.08$ & \\
& 4 & $0.40 \pm 0.02$ & $0.83 \pm 0.04$ & $20.44 \pm 1.24$ &$22.77 \pm 1.34$& $0.07 \pm 0.01$ & \\
& 5 & $1.06 \pm 0.03$ & $0.64 \pm 0.08$ & $5.55 \pm 0.99$ & $7.30 \pm 1.11$ & $-0.26 \pm 0.06$ & \\
& 6 & $6.54 \pm 0.64$ & $5.94 \pm 1.09$ & $2.95 \pm 0.69$ & $7.04 \pm 1.29$ & $0.07 \pm 0.09$& \\
& 7 & $1.91 \pm 0.02$ & $0.35 \pm 0.05$ & $3.13 \pm 0.45$ & $2.04 \pm 0.81$ & $0.10 \pm 0.24$ & \\
\midrule
\multirow{7}{*}{8} %
& 1 & $2.58 \pm 0.19$ & $2.55 \pm 0.41$ & $9.72 \pm 1.10$ & $14.05 \pm 2.20$ & $0.20 \pm 0.04$ & \multirow{7}{*}{397.55/418}\\
& 2 & 0.003(0.045)  & $0.36 \pm 0.13$ & $13.02 \pm 3.24$ & $11.93 \pm 2.41$ & $-0.06 \pm 0.03$ & \\
& 3$*$ & $1.55 \pm 0.05$ & $0.75 \pm 0.20$ & $5.72 \pm 1.72$ & $6.22 \pm 1.95$ & $0.47 \pm 0.13$ & \\
& 4 & $0.30 \pm 0.02$ & $0.59 \pm 0.11$ & $19.99 \pm 3.28$ & $19.03 \pm 3.09$ &$0.10 \pm 0.02 $& \\
& 5 & $0.66 \pm 0.11$ & $0.95 \pm 0.15$ & $12.05 \pm 3.25$ & $12.64 \pm 3.21$ & $-0.06 \pm 0.06$ & \\
& 6 & $4.53 \pm 0.81$ & $6.06 \pm 0.53$ & $4.54 \pm 1.51$ & $10.212 \pm 2.09$ & $0.21 \pm 0.06$& \\
& 7 & $2.01 \pm 0.08$ & $0.65 \pm 0.38$ & $3.23 \pm 2.1$ & 2.89(6.05) & $0.55 \pm 0.17$ & \\

\midrule
\multirow{5}{*}{9}

& 1 & $1.35 \pm 0.08$ & $2.84 \pm 0.07$& $13.58 \pm 0.52$ & $16.40 \pm 0.0.72$ & $0.23 \pm 0.01$& \multirow{5}{*}{430.65/430}\\
& 2 & $0.03 \pm 0.03$ & $0.66 \pm 0.01$ & $27.34 \pm 3.55$ & $19.77 \pm 4.06$ & $-0.06 \pm 0.07$ & \\
& 3$*$ & $1.11 \pm 0.03$ & $0.57 \pm 0.17$& $4.06 \pm 1.30$ & $3.52 \pm 1.70$ & $0.59 \pm 0.19$ & \\
& 4 & $0.19 \pm 0.04$ & $0.57 \pm 0.03$ & $14.04 \pm 7.20$ & $13.57 \pm 3.19$ & $0.33 \pm 0.19$& \\
& 5 & $3.18 \pm 2.00$ & $9.74 \pm 1.57$ & 0.001(2.04) & $9.18 \pm 1.21$ & $0.33 \pm 0.13$ &  \\
\bottomrule
\end{tabular}

\end{adjustbox}
\begin{tablenotes}[flushleft]
\footnotesize
\item \textit{Note.} The Value in parenthesis denotes the 84\% upper limits for the corresponding parameter. The symbol $*$ indicates the imaginary QPO component.
\end{tablenotes}
\end{table*}

\section{Phase lags and coherence function between 2--4~keV and 4--12~keV energy bands}
\label{cliff-dip_above2}
\begin{figure*}[t]
    \centering
    \includegraphics[scale=0.35]{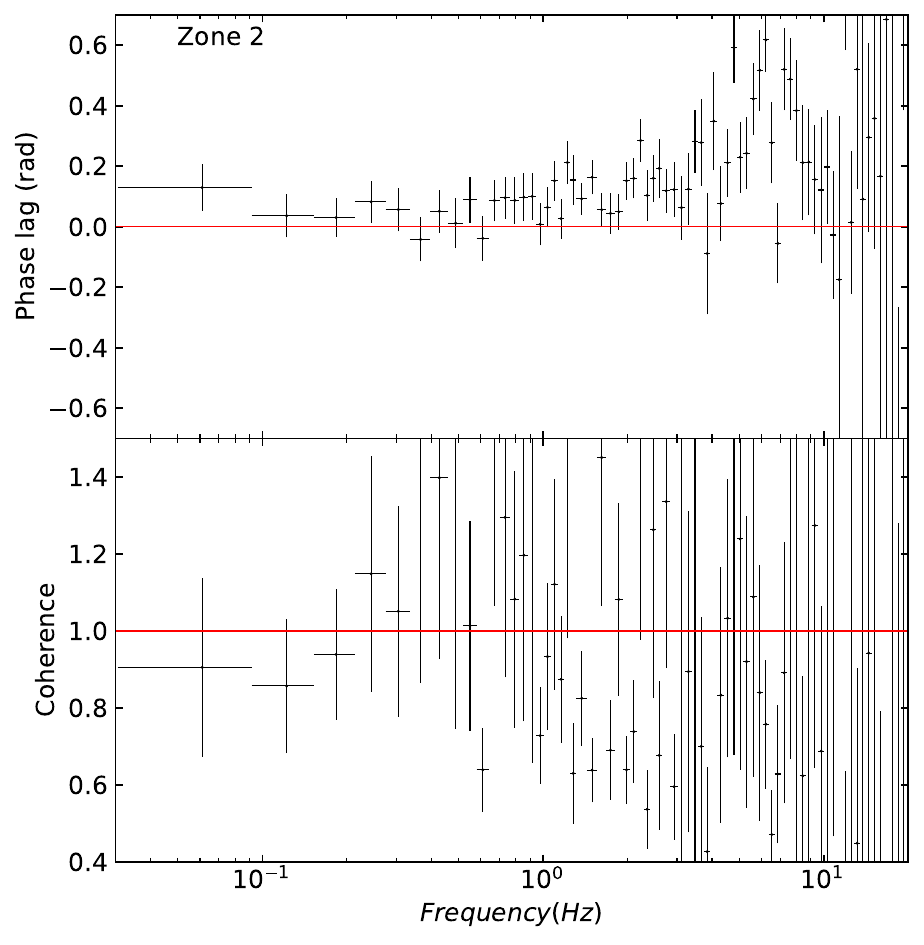}
    \includegraphics[scale=0.35]{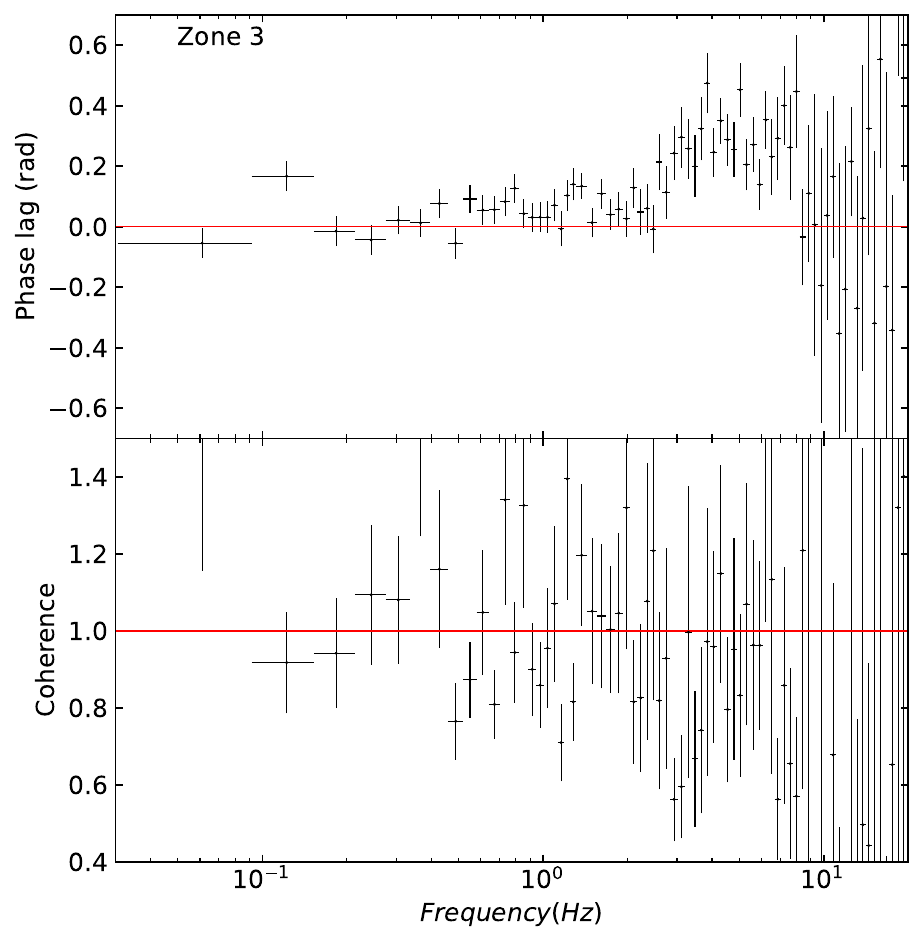}
    \includegraphics[scale=0.35]{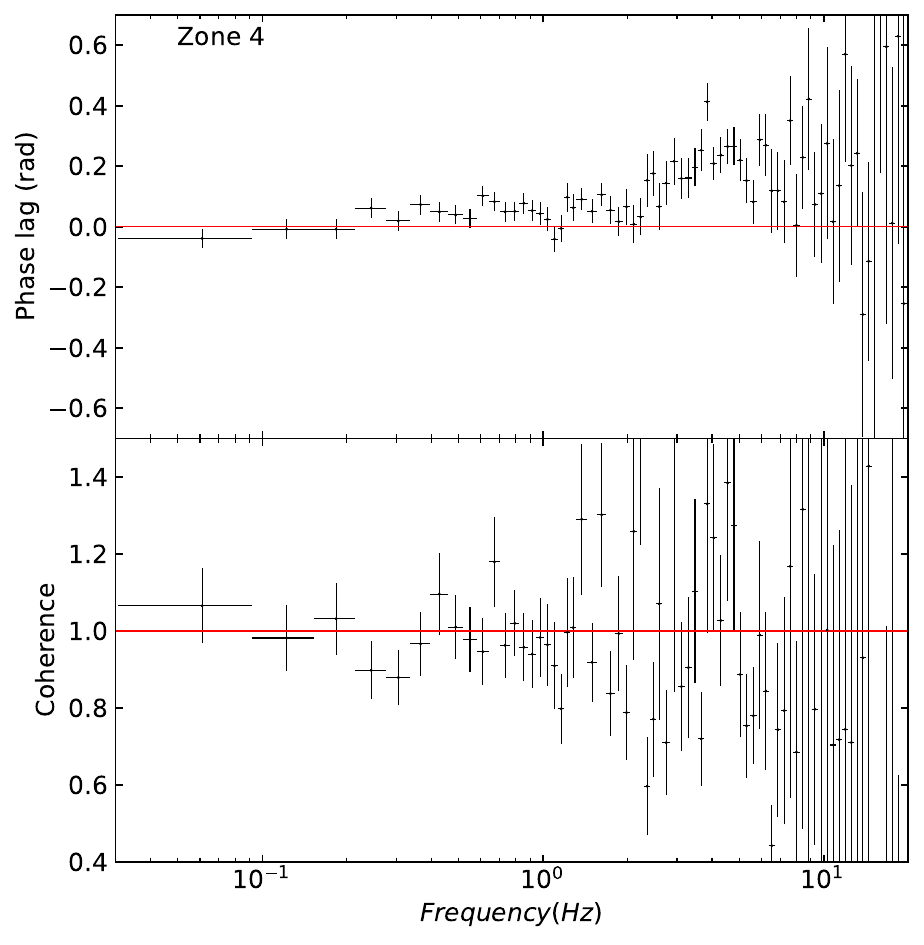}
    \includegraphics[scale=0.35]{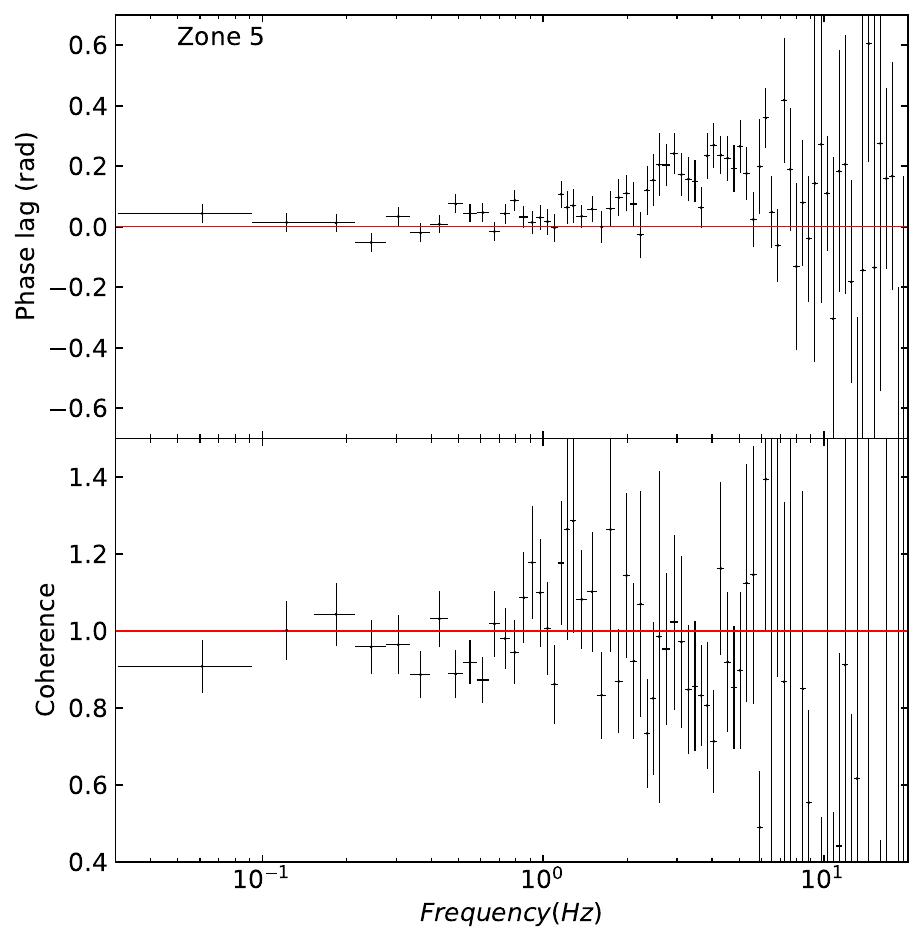}
    \includegraphics[scale=0.35]{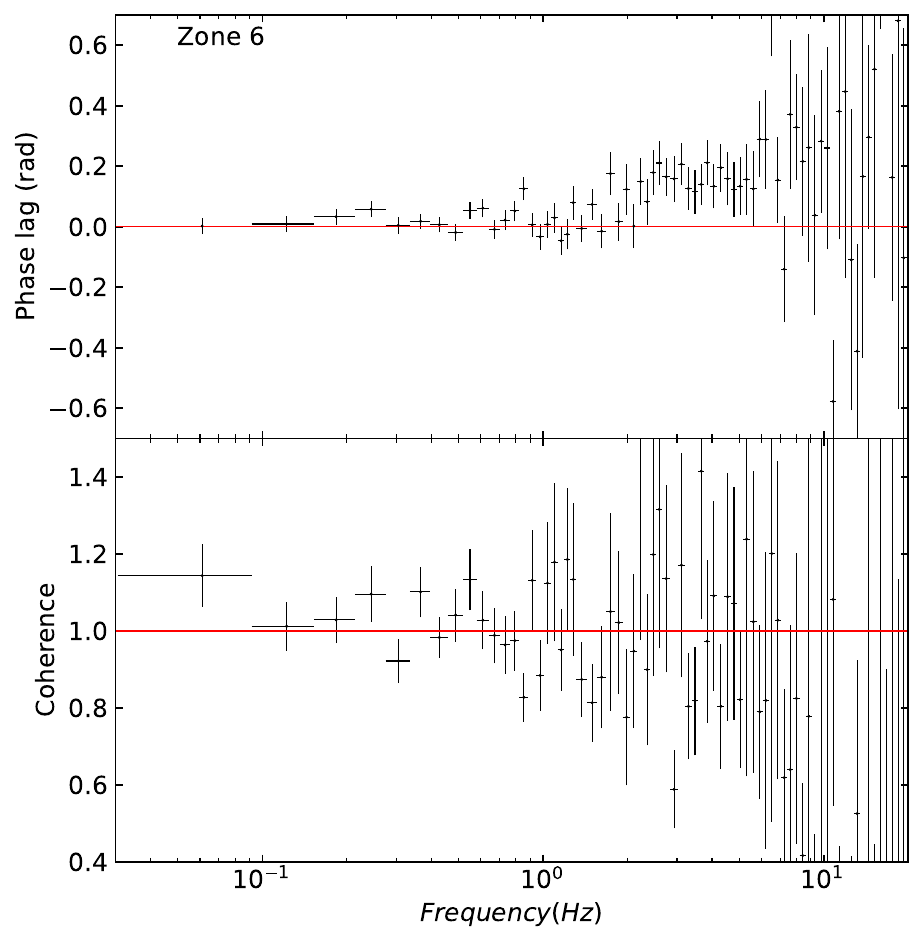}
    \caption{The phase lags and coherence function between the $2-4$~keV and $4-12$~keV energy bands. Note: we rebin the frequency spectrum by an factor of $10^{1/50}$ to improve the SNR.}
    \label{fig:cliff-dip_above2}
\end{figure*}

\citet{2024A&A...687A.284K} first reported that in Cygnus~X--1 the increase in the phase lags and the drop in the coherence usually disappear when the reference band is above $\sim$~2~keV \citep[see also][]{2025A&A...696A.237F}. The same behaviour was also observed in MAXI~J1820+070 \citep[][]{2025A&A...696A.128B}. 
Fig.~\ref{fig:cliff-dip_above2} shows the phase-lag frequency spectrum and coherence function of \so\ when we use a reference band above 2~keV. 
In Zones 2–6, an increase in the phase lags is apparent, whereas the corresponding drop in the coherence function is not clearly discernible. 
We also apply the multi-Lorentzian model to fit simultaneously the PDS in the $2-4$~keV and $4-12$~keV bands as well as the corresponding CS. 
In Zones 2--6, we can fit both the PDS and CS with a model consisting of four Lorentzian components. The phase-lag frequency spectra can be well described by the model derived from the corresponding best-fit model for both the PDS and CS. We find that a high-frequency broad component with a large imaginary part is responsible for the increase of the phase lags. In Zone~3, where the cliff–dip pair is the most apparent, adding a Lorentzian component at the possible dip frequency to both the PDS and CS reproduces the dip in the coherence function, although the component remains very weak and only marginally significant.

\end{appendix}

\end{document}